\def\fig{{\rm\thinspace Figure}}

\def\etal{{\it et al.\thinspace}}

\def\ie{{\it i.e.\ }}

\def\gcm3{{g cm${}^{-3}$}}

\def\msol{\hbox{$\rm\thinspace M_{\odot}\; $}}

\def\h50{\hbox{$\rm\thinspace h_{50}$}}
\def\h50m1{\hbox{$\rm\thinspace h_{50}^{-1}$}}

\def\etal{{\it et al.\thinspace}}

\def\ie{{\it i.e.\ }}

\def\fig{figure}

\def\p3m{P${}^3$M}
\def\ap3m{AP${}^3$M}
\def\-{{\em{---}}}
\def\Hydra{{\small HYDRA} }

\def\do{{\tt DO }}

\def\newspacing#1{%
  \def\baselinestretch{#1}\ifx\@currsize\normalsize %
  \@normalsize \else \@currsize\fi%
}

\documentclass{elsart}

\usepackage{natbib}
\usepackage{psfig}
\begin{document}

\runauthor{Thacker and Couchman}
\begin{frontmatter}
\title{A Parallel Adaptive P${}^3$M code with
Hierarchical Particle Reordering}

\author[Queens]{Robert J. Thacker\thanksref{CITA}}
\and
\author[McMaster]{H. M. P. Couchman}

\address[Queens]{Department of Physics, Engineering Physics and 
Astronomy, Queen's University, Kingston, 
Ontario, K7L 3N6}
\address[McMaster]{Department of Physics and Astronomy,
McMaster University, 1280 Main St. West, Hamilton, Ontario, L8S 4M1,
Canada.}
\thanks[CITA]{CITA National Fellow}
\begin{abstract}

We discuss the design and implementation of HYDRA\_OMP a parallel
implementation of the Smoothed Particle Hydrodynamics--Adaptive \p3m (SPH-A\p3m) code HYDRA. The code is designed primarily for
conducting cosmological hydrodynamic simulations and is written in
Fortran77+OpenMP. A number of optimizations for RISC processors and
SMP-NUMA architectures have been implemented, the most important
optimization being hierarchical reordering of particles within chaining
cells, which greatly improves data locality thereby removing the cache
misses typically associated with linked lists. Parallel scaling is good,
with a minimum parallel scaling of 73\% achieved on 32 nodes for a variety of modern SMP architectures. We give
performance data in terms of the number of particle updates per second,
which is a more useful performance metric than raw MFlops. 
A basic version of the code will be made available to the community in the near
future.\\

 \end{abstract}

\begin{keyword}
Simulation, cosmology, hydrodynamics, gravitation, structure formation\\
\PACS{02:60.-x 95:30.Sf 95:30.Lz 98:80.-k}
\end{keyword}

\end{frontmatter}
\section{Introduction} 

The growth of cosmological structure in the Universe is determined
primarily by (Newtonian) gravitational forces. Unlike the electrostatic
force, which can be both attractive and repulsive and for which
shielding is important, the ubiquitous attraction of the gravitational
force leads to extremely dense structures, relative to the average
density in the Universe. Galaxies, for example, are typically $10^4$
times more dense than their surrounding environment, and substructure
within them can be orders of magnitude more dense.  Modelling such large
density contrasts is difficult with fixed grid methods and,
consequently, particle-based solvers are an indispensable tool for
conducting simulations of the growth of cosmological structure. The
Lagrangian nature of particle codes makes them inherently adaptive
without requiring the complexity associated with adaptive Eulerian
methods. The Lagrangian Smoothed Particle Hydrodynamics
(SPH,\cite{GM77}) method also integrates well with gravitational solvers
using particles, and because of its simplicity, robustness and ability
to easily model complex geometries, has become widely used in cosmology.
Further, the necessity to model systems in which orbit crossing, or
phase wrapping, occurs (either in collisionless fluids or in collisional
systems) demands a fully Lagrangian method that tracks mass. 
While full
six-dimensional (Boltzmann) phase-space models have been attempted,
the resolution is still severely limited on current computers for most
applications. 

Particle
solvers of interest in cosmology can broadly be divided into hybrid
direct plus grid-based solvers such as Particle-Particle, Particle-Mesh
methods (\p3m,\cite{He81}) and ``Tree'' methods which use truncated low
order multipole expansions to evaluate the force from distant particles
\cite{BH86}. Full multipole methods \cite{GR87}, are slowly gaining
popularity but have yet to gain widespread acceptance in the
cosmological simulation community. There are also a number of hybrid
tree plus particle-mesh methods in which an efficient grid-based solver
is used for long-range gravitational interactions with sub-grid forces
being computed using a tree. Special purpose hardware \cite{Su90} has
rendered the direct PP method competitive in small simulations (fewer
than 16 million particles), but it remains unlikely that it will ever be
competitive for larger simulations.

The \p3m algorithm has been utilized extensively in cosmology. The first
high resolution simulations of structure formation were conducted by
Efstathiou \& Eastwood \cite{EE81} using a modified \p3m plasma code. In
1998 the Virgo Consortium used a \p3m code to conduct the first billion
particle simulation of cosmological structure formation \cite{Ev02}.
The well-known  problem of slow-down under heavy
particle clustering, due to a rapid rise in the number of short-range 
interactions, can be largely solved by the use of adaptive,
hierarchical, sub-grids \cite{C91}. Only when a regime is approached where
multiple time steps are beneficial does the adaptive \p3m (\ap3m)
algorithm become less competitive than modern tree-based solvers.
Further, we note that
a straightforward multiple time-step scheme has been implemented in 
\ap3m with a factor of 3 speed-up reported \cite{DE93}.

\p3m has also been vectorized by a number of groups including Summers \cite{SU93}.  
Shortly after, both Ferrell \& Bertschinger \cite{FB94} and Theuns
\cite{T94} adapted \p3m to the massively parallel architecture of the
Connection Machine. This early work highlighted the need for careful
examination of the parallelization strategy because of the load imbalance
that can result in gravitational simulations as particle clustering develops. Parallel versions of \p3m
that use a 1-dimensional domain decomposition, such as the P4M code of
Brieu \& Evrard \cite{BE00} develop large load imbalances under clustering
rendering them useful only for very homogeneous simulations.
Development of vectorized treecodes \cite{H90,HK89} predates the early 
work on \p3m codes 
and a discussion of a combined TREE+SPH (TREESPH) code for massively
parallel architectures is presented by Dav\'{e} \etal\, \cite{D97}. 
There are now a number of combined parallel TREE+SPH solvers 
\cite{W03,K03,V01,L00} and TREE gravity solvers \cite{MM00,UA01,D96}.
Pearce
\& Couchman \cite{PC97} have discussed the parallelization of \ap3m+SPH on
the Cray T3D using Cray Adaptive Fortran (CRAFT), which is a
directive-based parallel programming methodology. This code was developed
from the serial HYDRA algorithm \cite{CT95} and much of our discussion in
this paper draws from this first parallelization of \ap3m+SPH.  A highly
efficient distributed memory parallel implementation of \p3m using the 
Cray 
SHMEM library has been
developed by MacFarland \etal\, \cite{M98}, and further developments of
this code include a translation to MPI-2, the addition of \ap3m
subroutines and the inclusion of an SPH solver \cite{T03}. 
Treecodes have also been
combined with grid methods to form the Tree-Particle-Mesh solver
\cite{Jim,BO00,W02,B02,D03,VS05}. The algorithm is 
somewhat less efficient than \ap3m in a fixed time-step regime, but 
its simplicity offers advantages when multiple time-steps are 
considered \cite{VS05}. Another 
interesting, and highly efficient N-body 
algorithm is the Adaptive Refinement Tree (ART) method \cite{KK97} which 
uses a short-range force correction that is calculated via a multi-grid 
solver on refined meshes.

There are a number of factors in cosmology that drive researchers towards
parallel computing. These factors can be divided into the desire to 
simulate with the highest possible resolution, and hence particle number, and also the need to 
complete simulations in the shortest possible time frame to enable rapid 
progress. The desire for high resolution comes from two areas. Firstly,
 simultaneously simulating the growth of structure on
the largest and smallest cosmological scales requires enormous mass
resolution (the ratio of mass scales between a supercluster and the
substructure in a galaxy is $>10^9$). This problem is fundamentally
related to the fact that in the currently favoured Cold Dark Matter \cite{B84} cosmology
structure grows in a hierarchical manner. 
A secondary desire for high resolution comes from simulations that are
performed to make statistical predictions. To ensure the lowest possible
sample variance the largest possible simulation volume is desired. 

For complex codes, typically containing tens of thousands of lines, the
effort in developing a code for distributed-memory machines, using an API
such as MPI \cite{MPI}, can be enormous. The complexity within such codes
arises from the subtle communication patterns that are disguised in serial
implementations. Indeed, as has been observed by the authors, development
of an efficient communication strategy for a distributed memory version of
the \p3m code has required substantially more code than the \p3m algorithm 
itself (see
\cite{M98}). This is primarily because hybrid, or multi-part solvers, of which \p3m
is a classic example, have data structures that require significantly
different data topologies for optimal load balance at different stages of 
the solution cycle. Clearly a globally
addressable work space renders parallelization a far simpler task in such
situations. It is also worth noting that due to time-step 
constraints and the scaling of the algorithm with the number of 
particles, doubling 
the 
linear 
resolution along an axis of a simulation
increases the computational work load by a factor larger than
20; further doubling would lead to a workload in excess of 400 times
greater.

The above considerations lead to the following observation: modern SMP
servers with their shared memory design and superb performance
characteristics are an excellent tool for conducting simulations requiring
significantly more computational power than that available from a
workstation. Although such servers can never compete with massively
parallel machines for the largest simulations, their ease of use and programming renders them highly
productive computing environments. 
The OpenMP (http://www.openmp.org)
API for shared-memory programming is simple to 
use and enables loop level 
parallelism by the insertion of pragmas 
within the source code. 
Other than their limited expansion capacity, the strongest argument
against purchasing an SMP server remains hardware cost. However, there
is a trade-off between science accomplishment and development time that
must be considered above hardware costs alone.  Typically, programming a
Beowulf-style cluster for challenging codes takes far longer and
requires a significantly greater monetary and personnel investment on a 
project-by-project basis.  Conversely, for
problems that can be efficiently and quickly parallelized on a
distributed memory architecture, SMP servers are not cost effective. The
bottom line remains that individual research groups must decide which 
platform is most appropriate.

The code that we
discuss in this paper neatly fills the niche between workstation
computations and massively parallel simulations.
There is also a class of simulation problems in cosmology that 
have
particularly poor parallel scaling, regardless of the simulation algorithm
used (the fiducial example is the modelling of single galaxies, see \cite{TC00}). This class of problems corresponds to particularly
inhomogeneous particle distributions that develop a large disparity in
particle-update timescales (some particles may be in extremely dense
regions, while others may be in very low density regions). Only a very
small number of particles\-insufficient to be distributed
effectively across multiple nodes\-will require a large number of
updates due to their small time-steps. For this type of
simulation the practical limit of scalability appears to be order 10 PEs.

The layout of the paper is as follows: in section 2 we review the physical
system being studied. This is followed by an extensive exposition of the
\p3m algorithm and the improvements that yield the \ap3m algorithm.  The
primary purpose of this section is to discuss some subtleties that
directly impact our parallelization strategy. At the same time we also
discuss the SPH method and highlight the similarities between the two
algorithms. Section 2 concludes with a discussion of the serial HYDRA
code. Section 3 begins with a short discussion of the memory hierarchy 
in
RISC (Reduced Instruction Set Computer) systems, and how eliminating
cache-misses and ensuring good cache reuse ensures optimal performance on
these machines. This is followed by a discussion of a number of code
optimizations for RISC CPUs that also lead to performance improvements on
shared memory parallel machines (primarily due to increased data
locality). In particular we discuss improvements in particle bookkeeping,
such as particle index reordering. While particle reordering might be
considered an expensive operation, since it involves a global sort, it
actually dramatically improves run time because of bottlenecks in the
memory hierarchy of RISC systems.  In section 4 we discuss in detail the
parallelization strategies adopted in HYDRA\_OMP. To help provide further
understanding we compare the serial and parallel call trees. In section 5
we consolidate material from sections 3 \& 4 by discussing considerations for NUMA
machines and in particular the issue of data placement.  Performance
figures are given in Section 6, and we present our conclusions in section
7.

\section{Review of the serial algorithm}
\subsection{Equation set to be solved}
The simulation of cosmic structure formation is posed as an initial value
problem. Given a set of
initial conditions, which are usually constrained by experimental data,
such as the WMAP data \cite{WM03}, we must solve the following
gravito-hydrodynamic equations;
\begin{enumerate}
\item the {continuity equations},
\begin{equation}
{ d \rho_g \over dt}+\rho_g \nabla. {\bf {v}}_g=0,\;\;\;{ d \rho_{dm} 
\over 
dt}+\rho_{dm}\nabla. {\bf {v}}_{dm}=0
\label{gravito1}
\end{equation}
where $g$ denotes gas and  $dm$ dark matter.
\item the Euler and acceleration equations,
\begin{equation}
{d {\bf {v}}_g \over dt}=-{1 \over \rho_g}
\nabla P-\nabla \phi, \;\;\;
{d {\bf {v}}_{dm} \over dt}=-\nabla \phi,
\end{equation}
\item the {Poisson equation},
\begin{equation}
\nabla^2 \phi = 4 \pi G (\rho_g+\rho_{dm}),
\label{poisson}
\end{equation}
\item the entropy conservation equation,
\begin{equation}
{ds \over dt}=0,
\end{equation}
\end{enumerate} 
where the conservation of entropy is a result of ignoring dissipation,
viscosity and thermal conductivity (\ie an ideal fluid). The dynamical
system is closed by the equation of state $P=P(\rho_g,s)$. We assume an 
ideal gas equation of state, with $\gamma=5/3$ in our code, although 
many others are possible.
Alternatively,
the entropy equation can be substituted with the
conservation of energy equation,
\begin{equation}
{d u \over dt} = -{P \over\rho_g} \nabla. {\bf v}_g,
\label{gravito2}
\end{equation}
and the equation of state is then $P=P(\rho_g,u)$. We note that the use 
of a particle-based method ensures that the continuity equations are 
immediately satisfied.

\subsection{Gravitational solver}
Let us first discuss the basic features of the \p3m algorithm, a
thorough review can be found in \cite{He81}. The
fundamental basis
of the \p3m algorithm is that the gravitational force can be separated
into short and long range components, \ie,
\begin{equation}
{\bf F}_{grav}={\bf F}_{short}+{\bf F}_{long},
\end{equation}
where ${\bf F}_{long}$ will be provided by a Fourier-based solver and
${\bf F}_{short}$ will be calculated by summing over particles within a
given short range radius. The ${\bf F}_{long}$ force is typical known as
the PM force, for Particle-Mesh, while the ${\bf F}_{short}$ range force
is typical known as the PP force, for Particle-Particle.
The accuracy of the ${\bf F}_{grav}$ force can be improved
by further smoothing the mesh force, ${\bf F}_{long}$, and hence increasing the range over the which the short-scale calculation is
done, at the cost of an increased number of particle--particle interactions.

The first step in evaluating that PM force is to interpolate the mass 
density of the particle 
distribution on to a grid which can be viewed as a map from a 
Lagrangian representation to an 
Eulerian one.  The
interpolation function we use is the the `Triangular Shaped Cloud' (TSC)
`assignment function' (see \cite{He81} for a detailed discussion of
possible assignment functions). Two benefits of using TSC are
good suppression of aliasing from power above the Nyquist frequency of
the grid and a comparatively low directional force error around 
the grid spacing.  The mass
assignment operation count is ${\bf O} (N)$, where $N$ is the number of
particles.

Once the mass density grid has been constructed it is Fourier transformed
using an FFT routine, which is an ${\bf O} (L^3 \log L)$ operation, where
$L$ is the extent of the Fourier grid in one direction.  The resulting
k-space field is then multiplied with a Green's function that is
calculated to minimize errors associated with the mass assignment
procedure (see Hockney \& Eastwood for a review of the `Q-minimization'
procedure). Following this convolution, the resulting potential grid is
differenced to recover the force grid. We use a 10-point differencing
operator which incorporates off-axis components and reduces directional force errors, but many others are possible. Finally, the PM accelerations are
found from the force grid using the mass assignment function to
interpolate the acceleration field. The PM algorithm has an operation
cost that is approximately ${\bf O} (\alpha N+\beta L^3 \log L)$ where $\alpha$ and
$\beta$ are constants (the ${\bf O} (L^3)$ cost of the differencing is 
adequately approximated by the logarithmic term describing the FFT) .

\begin{figure}[t]
\vspace{10cm}
\includegraphics{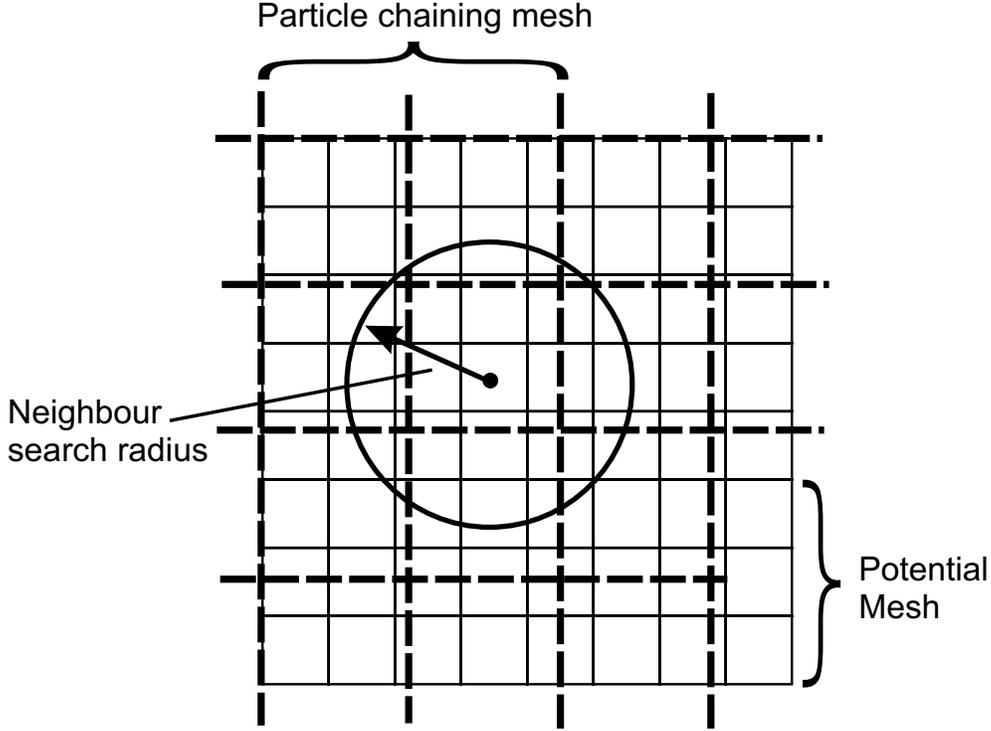}
\caption{Overlay of the chaining mesh on top of the potential mesh to
show spacing and the search radius of the short range force. These two 
meshes do not need to be commensurate except on the scale of the box, the 
required matching of forces is achieved by shaping the long and short 
range force components, ${\bf F}_{short}$ and ${\bf F}_{long}$.} 
\label{chaining} \end{figure}

Resolution above the Nyquist frequency of the PM code, or equivalently 
sub PM grid resolution, is provided by the 
pair-wise (shaped) short-range force summation.
Supplementing the PM force with the short-range PP force gives
the full
P${}^3$M algorithm, and the execution time scales approximately in
proportion to $\alpha$N+$\beta$L${}^3$ log L + $\gamma \sum $N${}^2_{pp}$,
where $\gamma$ is a constant and N${}^2_{pp}$ corresponds to the number of
particles in the short range force calculation within a specified region.
The summation is performed over all the PP regions, which are identified
using a chaining mesh of size {\em Ls}${}^3$; see \fig~\ref{chaining} 
for an illustration of the
chaining mesh overlaid on the potential mesh. P${}^3$M suffers the
drawback that under heavy gravitational clustering the short range sum
used to supplement the PM force slows the calculation down dramatically -
the N${}^2_{pp}$ term dominates as an increasingly large number of 
particles contribute to the
short range sum. Although acutely dependent upon the particle number and
relative clustering in a simulation, the algorithm may slow down by a
factor between 10-100 or possibly more. While finer meshes 
partially alleviate 
this problem they quickly become inefficient due to wasting computation 
on areas that do not need higher resolution.

Adaptive P${}^3$M remedies the slow-down under clustering of \p3m by
isolating regions where the N${}_{pp}^2$ term dominates and solving for
the short range force in these regions using FFT methods on a sub-grid,
which is then supplemented by short range calculations involving fewer
neighbours.  This process is a repeat of the \p3m algorithm on the
selected regions, with an isolated FFT and shaped force. 
At the expense of a 
little additional bookkeeping,
this method circumvents the sometimes dramatic slow-down of \p3m.  
The operation count
is now approximately, 
\begin{equation} 
\alpha N + \beta L^3 \log L +
\sum_{j=1}^{n_{ref}} \left[ \alpha_j N_j + \beta_j L_j^3 \log L + \gamma_j 
\sum
N_{j_{pp}}^2 \right], 
\end{equation} where $n_{ref}$ is the number of
refinements. The $\alpha_j$ and $\gamma_j$ are all expected to be very
similar to the $\alpha$, and $\gamma$ of the main solver, while the
$\beta_j$ are approximately four times larger than $\beta$ due to the
isolated Fourier transform. 
Ideally during the course of 
the simulation the time per iteration approaches a constant, roughly 2-4 
times that of a uniform distribution (although when the SPH algorithm is 
included this slow-down can be larger).

\subsection{SPH solver}
When implemented in an adaptive form \cite{W81}, with smoothing performed over a fixed number of neighbour particles, SPH is an order $N$ 
scheme and fits
well within the P${}^3$M method since
the short-range force-supplement for the mesh force can be used  to find 
the particles which are required for the SPH calculation. There
are a number of excellent reviews of the SPH methodology
\cite{HK89,M92,S96} and we present, here, only those details necessary to understand our specific algorithm implementation. Full details of our implementation can be found in \cite{rob2}.

We use an explicit `gather' smoothing kernel and the symmetrization of the
equation of motion is achieved by making the replacement,
\begin{equation}
\nabla_j \overline{W}({\bf r}_i-{\bf r}_j,h_j,h_i) = -
\nabla_i \overline{W}({\bf r}_i-{\bf r}_j,h_i,h_j) + {\bf O}(\nabla h)
\end{equation}
in the `standard' SPH equation of motion (see \cite{S96}, for
example). Note that the sole purpose of `kernel averaging' in this
implementation, denoted by
the bar on the smoothing kernel $W$, is to ensure that the above 
replacement is correct to ${\bf
O}(h)$. Hence the equation of motion is,
\[
{d {\bf v}_i \over dt}=
- \sum_{j=1,r_{ij}<2h_i}^N m_j \; ({P_i \over \rho_i^2}+{\Pi_{ij} \over
2}) \;
\nabla_i
\overline{W}({\bf r}_i-{\bf r}_j,h_i,h_j)
\]
\begin{equation}
\;\;\;\;\;\;\;\;\;\;\;\;\; +
\sum_{j=1,r_{ij}<2h_j}^N
m_j \; ({P_j \over \rho_j^2}+{\Pi_{ji} \over 2}) \; \nabla_j
\overline{W}({\bf r}_i-{\bf r}_j,h_j,h_i).
\end{equation}
The artificial viscosity, $\Pi_{ij}$, is used to prevent interpenetration
of particle flows and is given by,
\begin{equation}
\Pi_{ij}={ -\alpha \mu_{ij} \bar{c}_{ij} + \beta \mu_{ij}^2 \over
\tilde{\rho}_{ij}}f_i,
\end{equation}
where,
\begin{equation}
\mu_{ij}=\cases{
 \bar{h}_{ij} {\bf v}_{ij}.{\bf r}_{ij} / (r_{ij}^2+\nu^2), &
${\bf v}_{ij}.{\bf r}_{ij}<0$;
\cr
0, & ${\bf v}_{ij}.{\bf r}_{ij} \geq 0$,
}
\end{equation}
\begin{equation}
\tilde{\rho}_{ij} = \rho_i(1+(h_i/h_j)^3)/2,
\end{equation}
and
\begin{equation}
f_i={|\!< \nabla {\bf. v}>_i\!\!| \over |\!<\nabla {\bf. v}>_i\!\!| + 
|\!<\nabla
{\bf \times} {\bf v}>_i\!\!| +
0.0001c_i/h_i }.
\end{equation}
with bars being used to indicate averages over the $i,j$ indices.
Shear-correction \cite{B95,NS97}, is
achieved by including the $f_i$ term which reduces the\-unwanted\- 
artificial viscosity in shearing flows. Note that the lack of
$i-j$ symmetry in $\Pi_{ij}$ is not a concern since the equation of motion
enforces force symmetry. 

The energy equation is given by,
\begin{equation}
{d u_i \over dt}=\sum_{j=1,r_{ij}<2h_i}^N
 m_j ({P_i \over
\rho^2_i}+{\Pi_{ij} \over 2}) \;
({\bf v}_i-{\bf v}_j).\nabla_i \overline{W}({\bf r}_i-{\bf r}_j,h_i,h_j).
\end{equation}

The solution of these equations is comparatively straightforward. As in
the \ap3m solver it is necessary to establish the neighbour particle
lists.  The density of each particle must be evaluated and then, in a
second loop, the solution to the force and energy equations can be found.
Since the equation of motion does not explicitly depend on the density of
particle $j$ (the artificial viscosity has also been constructed to avoid
this) we emphasize that there is no need to calculate all the density
values first and then calculate the force and energy equations. If one
does calculate all densities first, then clearly the list of neighbours is
calculated twice, or alternatively, a large amount of memory must be used
to store the neighbour lists of all particles.  Using our method the
density can be calculated, one list of neighbours stored, and then the
force and energy calculations can be quickly solved using the stored list
of neighbours (see \cite{CT95}).


\subsection{Summary of solution cycle for each iteration}
As emphasized, the list data-structure used in the short-range force
calculation provides a common feature between the \ap3m and SPH solvers.
Hence, once a list of particle neighbours has been found, it is simple 
to
sort
through this and establish which particles are to be considered for
the gravitational calculation and the SPH calculation.
Thus the incorporation of SPH into AP${}^3$M necessitates only the
coordination of scalings and minor bookkeeping. The   
combined adaptive P${}^3$M-SPH code, `{\small HYDRA}', in serial FORTRAN
77
form is available on the World Wide Web from
http://coho.physics.mcmaster.ca/hydra.

The solution cycle of one time-step may be summarized as follows,
\begin{enumerate}
\item Assign mass to the Fourier mesh.
\item Convolve with the Green's function using the FFT method to get   
potential. Difference this to recover mesh forces in each dimension.  
\item Apply mesh force and accelerate particles.
\item Decide where it is more computationally efficient to solve via
the further use of Fourier methods as opposed to short-range forces and,
if so, place a new sub-mesh (refinement) there.
\item Accumulate the gas forces (and state changes) as well as the short
range gravity for all positions not in sub-meshes.
\item Repeat 1-5 on all sub-meshes until forces on all particles in
simulation have been accumulated.
\item Update time-step and repeat
\end{enumerate}

Note that the procedure of placing meshes is hierarchical in that a
further sub-mesh may be placed inside a sub-mesh. This procedure can
continue to an arbitrary depth but, typically, even for the most clustered
simulations, speed-up only occurs to a depth of six levels of refinement.

\begin{figure}[t]
\vspace{7cm}
\includegraphics{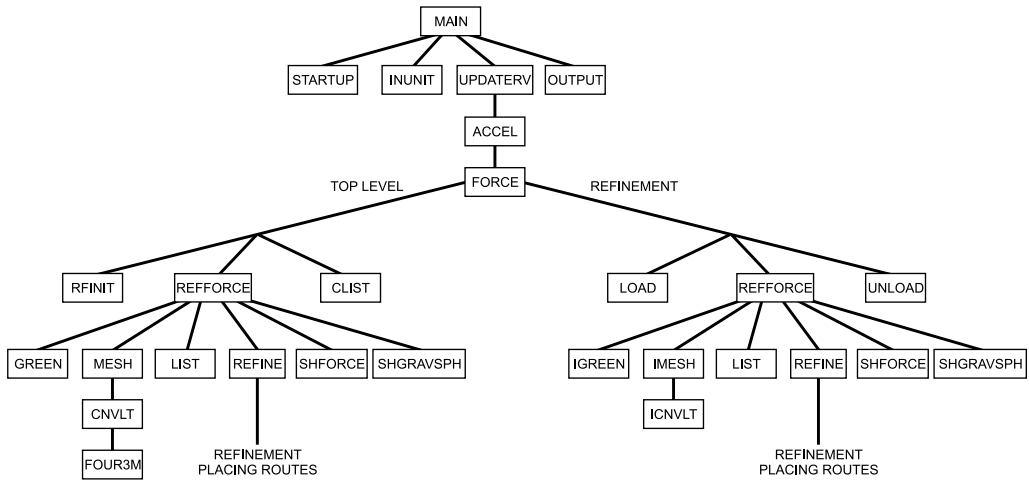}
\caption{Call tree of the HYDRA serial algorithm. Only significant
subroutines are shown for clarity. The refinement routines are the same as
the top level routines, modulo the lack of periodic wrap-around. In an
object-oriented framework these routines would be prime candidates for
overloading.}
 \label{ctree} 
\end{figure}

A pseudo call-tree for the serial algorithm can be seen in
\fig~\ref{ctree}. The purpose of each subroutine is as follows,
\begin{itemize}
\item{STARTUP} Reads in data and parameter files
\item{INUNIT} Calculates units of simulation from parameters in start-up
files
\item{UPDATERV} Time-stepping control
\item{OUTPUT} Check-pointing and scheduled data output routines
\item{ACCEL} Selection of time-step criteria and corrections, if
necessary, for comoving versus physical coordinates
\item{FORCE} Main control routine of the force evaluation subroutines
\item{RFINIT \& LOAD} Set up parameters for PM and PP calculation, in LOAD
data is also loaded into particle buffers for the refinement.
\item{CLIST \& ULOAD} Preparation of particle data for any refinements
that may have been placed, ULOAD also unloads particle data from
refinement buffers
\item{REFFORCE} Call PM routines, controls particle bookkeeping, call PP
routines.
\item{GREEN \& IGREEN} Calculation of Green's functions for periodic
(GREEN) and isolated (IGREEN) convolutions.
\item{MESH \& IMESH} Mass assignment, convolution call, and calculation of
PM acceleration in the periodic (MESH) and isolated (IMESH) solvers.
\item{CNVLT \& ICNVLT} Green's function convolution routines.
\item{FOUR3M} 3 dimensional FFT routine for periodic boundary conditions.
\item{LIST} Evaluation of chaining cell particle lists
\item{REFINE} Check whether refinements need to be placed.
\item{SHFORCE} Calculate force look-up tables for PP
\item{SHGRAVSPH} Evaluate PP and SPH forces      
\end{itemize}

\section{Optimizations of the serial code for RISC processors}
\subsection{Memory hierarchy of the RISC architecture}
The architecture of RISC CPUs incorporates a memory hierarchy with 
widely
differing levels of performance. Consequently, the efficiency of a code
running on a RISC processor is dictated almost entirely by the ratio of
the time spent in memory accesses to the time spent performing
computation. This fact can lead to enormous differences in code
performance.

The relative access times for the hierarchy are almost logarithmic.  
Access to the first level of cache memory takes 1-2 processor cycles,
while access to the second level of cache memory takes approximately 5
times as long. Access to main memory takes approximately 10 times longer.
It is interesting to note that SMP-NUMA servers provide further levels to
this hierarchy, as will be discussed later.

To improve memory performance, when retrieving a word from main 
memory three other words are typically retrieved: the `cache line'.
If the additional words are used within the computation on a short time 
scale, the algorithm exhibits good cache reuse. It is also important to 
not access memory in disordered fashion, \ie optimally one should need 
memory references that are stored within caches. Thus to exhibit 
good performance on a RISC processor, a code 
must exhibit both good cache reuse and a low number of cache misses. In
practice, keeping cache misses to a minimum is the first objective since
cache reuse is comparatively easy to achieve given a sensible ordering of
the calculation (such as a FORTRAN {\tt DO} loop).

\subsection{Serial Optimizations}
A number of optimizations for particle codes that run on RISC processors
are discussed in Decyk \etal \cite{D96}. Almost all of these
optimizations are included within our serial code, with the exception of
the mass assignment optimizations. Indeed a large number of their
optimizations, especially those relating to combining x, y, z coordinate
arrays into one 3-d array, can be viewed as good programming style.
While Decyk \etal\, demonstrate that the complexity of the periodic mass
assignment function prevents compilers from software pipelining the mesh
writes, we do not include their suggested optimization of removing the
modulo statements and using a larger grid. However, the optimization is
naturally incorporated in our isolated solver.

The first optimization we attempted was the removal of a `vectorizeable'
Numerical Recipes FFT used within the code (FOURN, see \cite{NR}).
Although the code uses an optimized 3-d FFT that can call the FOURN routine
repeatedly using either 1-d or 2-d FFT strategy (to reduce the number of
cache misses exhibited by the FOURN routine when run in 3-d), the overall
performance remains quite poor. Therefore we replaced this routine with
the FFTPack (see \cite{S82}) routines available from Netlib, and
explicitly made the 3-d FFT a combination of 1-d FFTs. Although there is
no question that FFTW \cite{FJ98} provides the fastest FFTs on almost all
architectures we have found little difference between FFTPack and FFTW
within our parallel 3-d FFT routine.  The greatest performance 
improvement is seen
in the isolated solver where the 3-d FFT is compacted to account for the
fact that multiple octants are initially zero.

Linked lists (hereafter the list array is denoted {\tt ll}) are a common
data structure used extensively in particle-in-cell type codes (see
\cite{He81}, for an extensive review of their use).  For a list of
particles which is cataloged according to cells in which they reside, it
is necessary to store an additional array which holds the label of the
first particle in the list for a particular cell. This array is denoted
{\tt ihc} for Integer Head of Chain. List traversal for a given cell is
frequently programmed in FORTRAN using an {\tt IF...THEN...GOTO}
structure (although it can be programmed with a {\tt DO WHILE} loop),
with the loop exiting on the {\tt IF} statement finding a value of zero
in the linked list.  Since the loop `index' (the particle index {\tt i})
is found recursively the compiler cannot make decisions about a number
of optimization processes, particularly software pipelining, for which
\do loops are usually better. Additionally, if the particles' indices
are not ordered in the list traversal direction then there will usually
be a cache miss in finding the element {\tt ll(i)} within the linked
list array.  Within the particle data arrays, the result of the particle
indices not being contiguous is another series of cache misses. Since a
number of arrays must be accessed to recover the particle data, the
problem is further compounded, and removal of the cache miss associated
with the particle indices should improve performance significantly.

\begin{figure}[t]
\vspace{90mm}
\includegraphics{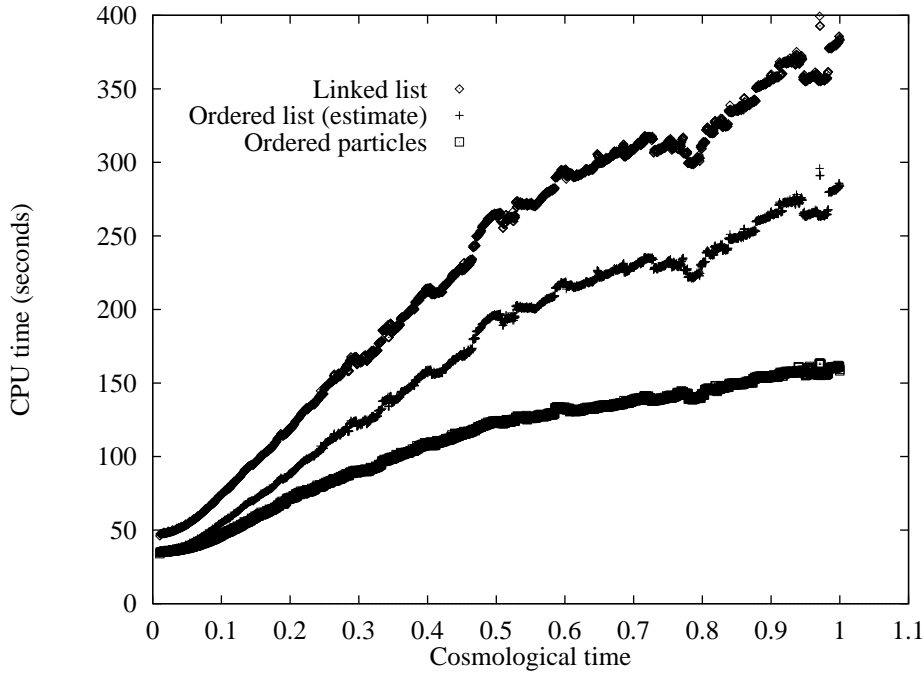}
\caption{Effect of changing the list structure on the execution time per 
iteration of 
the entire algorithm. We show results for the standard linked list 
implementation, ordered list and ordered particles. The ordered list 
times were estimated by taking the ratio between the linked list time and 
the ordered list time at t=1 and scaling the rest of the linked list 
values by this factor. The simulation was the Santa Barbara galaxy 
cluster simulation \cite{SB99} and it was conducted 
on a 266 Mhz Pentium III PC.}
 \label{timings}
\end{figure}

The first step that may be taken to improve the situation is to remove the
cache misses associated with the searching through the linked list. To do
this the list must be formed so that it is ordered. In other words the
first particle in cell {\tt j}, is given by {\tt ihc(j)}, the second
particle is given by {\tt ll(ihc(j))}, the third by {\tt ll(ihc(j)+1)}
{\em et cetera}.  This ordered list also allows the short range force
calculation to be programmed more elegantly since the {\tt IF..THEN..GOTO}
structure of the linked list can be replaced by a \do loop. However, since
there remains no guarantee that the particle indices will be ordered, 
the compiler is still heavily constrained in terms of the optimizations 
it may attempt, but the situation is distinctly better than for the 
standard linked list.
Tests performed on
this { ordered list} algorithm show that a 30\% improvement in speed is
gained over the linked list code (see \fig~\ref{timings}). Cache misses 
in the data arrays are of course still present in this algorithm.

As has been discussed, minimizing cache misses in the particle data 
arrays requires accessing them with a contiguous index. This means that 
within a given chaining cell the particle indices must be contiguous.
This can 
be achieved by reordering the indices of particles within chaining cells 
at each step of the iteration
(although if particles need to be tracked a permutation array must be 
carried). 
This {\em particle reordering} idea was realized comparatively early 
and has been discussed in the literature \cite{AS95,D96,rob1,M98}. A 
similar concept has been applied by Springel \cite{VS05} who uses
Peano-Hilbert ordering of particle indices to ensure data locality.  
However, in \p3m codes, prior to the implementation presented here only 
Macfarland
\etal \cite{M98} and Anderson and Shumaker \cite{AS95}, actually revised
the code to remove linked lists, other codes simply reordered the 
particles every few steps to reduce the probability of cache misses and achieved a 
performance improvement of up to 45\% \cite{M98}. 
Since the
adaptive refinements in \Hydra use the same particle indexing method, 
the particle
ordering must be done within the data loaded into a refinement, \ie
hierarchical rearrangement of indices results from the use of 
refinements.

The step-to-step permutation is straightforward to calculate: 
first the
particle indices are sorted according to their z-coordinate and then
particle array indices are simply changed accordingly.  
It is important to note that this method of particle
bookkeeping removes the need for an index list of the particles
(although in practice this storage is taken by the permutation array).  
All that need be stored is the particle
index corresponding to the first particle in the cell and the number of
particles in the cell.  On a RISC system particle reordering is so
efficient that the speed of the \Hydra simulation algorithm {\em more
than doubled}. For example, at the end of the Santa Barbara galaxy
cluster simulation, the execution time was reduced from 380 seconds to
160 seconds on a 266 Mhz Pentium III processor. On a more modern 2 
Ghz AMD Opteron, which has four times the L2 cache of a Pentium III, 
considerably better prefetch, as well as an on-die memory controller to 
reduce latency, 
we found the performance improvement for the final  
iterations to be a reduction in time from 29 seconds to 17.
This corresponds to a speed improvement of a factor of 1.7, which, while 
slightly less impressive than the factor of 2.4 seen on the older 
Pentium III, is still a significant improvement.
A comparison plot of the
performance of a linked list, ordered list and ordered particle code is
shown in \fig~\ref{timings}.

\section{Parallel Strategy}
Particle-grid codes, of the kind used in cosmology, are difficult to   
parallelize efficiently. The
fundamental limitation to the code is the degree to which the problem may
be subdivided while still averting race conditions and unnecessary 
buffering or synchronization. For example, the fundamental limit on the 
size of 
a computational atom in the PP code is effectively a chaining cell, 
while for 
the FFT routine it is a plane in the data cube. In practice, load 
balance constraints come into play earlier than theoretical limits as 
the work within the minimal atoms will rarely be equal (and can be 
orders of magnitude different).
Clearly these considerations set an upper bound on the degree to 
which the 
problem
can be subdivided, which in turn limits the number of processors that 
may
be used effectively for a given problem size. The code is a good example
of Gustafson's conjecture: a greater degree of parallelism may not allow
arbitrarily increased execution speed for problems of fixed size, but
should permit larger problems to be addressed in a similar time.

At an abstract level, the code divides into essentially two pieces: the 
top level mesh and 
the refinements. Parallelization of the top level mesh involves 
parallelizing the work in each associated subroutine.
Since an individual refinement may have very little work a    
parallel scheme that seeks to divide work at all points during execution
will be highly inefficient. Therefore the following division of
parallelism was made: conduct all refinements of size greater than $N_r$
particles across the whole machine, for refinements with less than $N_r$
particles use a list of all refinements and distribute one refinement to
each processor (or thread) in a task farm arrangement. On the T3D the limiting $N_r$ was 
found to
be approximately 32,768 particles, while on more modern machines we have  
found that 262,144 is a better limit. 

\begin{figure}[t]
\vspace{100mm}
\includegraphics{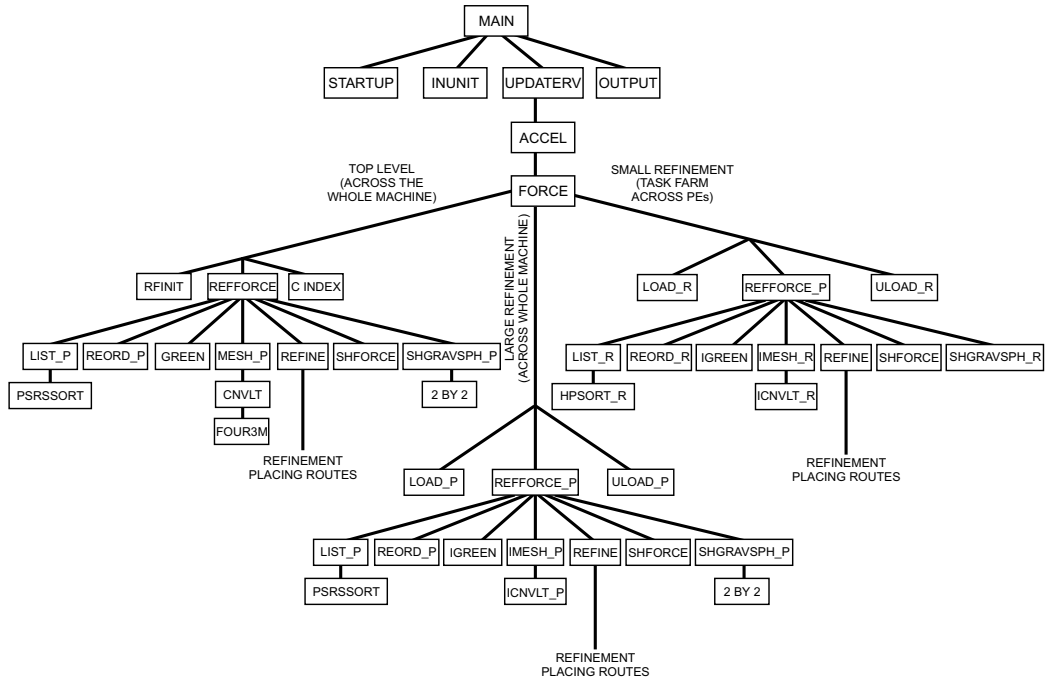}
\caption{Call tree of the HYDRA\_OMP algorithm. Only significant
subroutines are shown for clarity. The call tree is similar to the serial
algorithm except that a new class of routines is included for large
refinements. Where possible conditional parallelism has been used to
enable the reuse of subroutines in serial or parallel.}
 \label{ptree}
\end{figure}

In the following discussion the term processor element (PE) is used to 
denote a parallel execution thread.
Since only one thread of
execution is allotted per processor (we do not attempt load balancing 
via parallel slackness), this number is equivalent to the
number
of CPUs, and the two terms are used interchangeably. The call tree of the
parallel algorithm is given in figure~\ref{ptree}.

\subsection{The OpenMP standard} 
The OpenMP API supports a number of parallel constructs, such as executing 
multiple serial regions of code in parallel (a single program 
multiple data model), as well as the more typical 
loop-based parallelism model (sometimes denoted `PAR DO's), where the 
entire set of
loop iterations is distributed across all the PEs. The pragma for 
executing a loop in parallel, {\tt C\$OMP PARALLEL DO}  is placed before 
the {\tt DO} loop within the code body. Specification statements are 
necessary to inform the compiler about which variables are loop `private' 
(each processor carries its own value) and `shared' variables. A full 
specification of the details for each loop takes only a few lines of code, 
preventing the `code bloat' often associated with distributed 
memory parallel codes. 

\subsection{Load balancing options provided by the OpenMP standard}  
We use loop level parallelism throughout our code. To optimize load 
balance in a given routine it is necessary to select the most optimal 
iteration scheduling algorithm.
The OpenMP directives allow for the
following types of iteration scheduling:

\begin{itemize}

\item static scheduling - the iterations are divided into chunks (the size
of which may be specified if desired) and the chunks are distributed
across the processor space in a contiguous fashion. A cyclic 
distribution, or a cyclic distribution of small chunks is also 
available.

\item dynamic scheduling - the iterations are again divided up into  
chunks, however as each processor finishes its allotted chunk, it
dynamically obtains the next set of iterations, via a master-worker 
mechanism.

\item guided scheduling - is similar to static scheduling except that the
chunk size decreases exponentially as each set of iterations is finished.
The minimum number of iterations to be allotted to each chunk may be
specified.

\item runtime scheduling - this option allows the decision on which
scheduling to use to be delayed until the program is run. The desired
scheduling is then chosen by setting an environment variable in the
operating system.

\end{itemize}
The \Hydra code uses both static and dynamic scheduling.

\subsection{Parallelization of  
particle
reordering and permutation array creation}
While the step-to-step permutation is in principle simple to calculate,
the creation of the list permutation array must be done carefully to 
avoid race
conditions. An effective strategy is to calculate 
the chaining cell residence for each particle and then sort into bins of 
like chaining cells. Once particles have been binned in this fashion the 
rearrangement according to z-coordinates is a local permutation among 
particles in the chaining cell. 
 Our parallel algorithm works as follows:
\begin{enumerate}
\item First calculate the chaining cell that each particle
resides in,
and store this in an array
\item Perform an increasing-order global sort over the array of box
indices
\item Using a loop over particle indices, find the first
particle in each section of contiguous like-indices (the {\tt ihc} 
array)
\item Use this array to establish the number of particles in each
contiguous section (the {\tt nhc} array)
\item Write the z-coordinates of each particle within the chaining cell
into another auxiliary array
\item Sort all the non-overlapping sublists of z-coordinates for all 
cells in 
parallel while at the same time permuting an index array to store the
precise rearrangement of particle indices required
\item Pass the newly calculated permutation array to a routine 
that will rearrange all the particle data into the new order
\end{enumerate}
The global sort is performed using parallel sorting by regular sampling, 
\cite{ll93} 
with a code developed in part by J. Crawford and C. Mobarry. This code 
has been demonstrated to scale extremely well on
shared-memory architectures provided the
number of elements per CPU exceeds 50,000. This is significantly less
than our ideal particle load per processor (see section 6). For the
sorts within cells, the slow step-to-step evolution of particle
positions ensures data rearrangement is sufficiently local for this
to be an efficient routine. Hence we expect
good
scaling for the sort routines at the level of granularity we
typically use.

\subsection{Parallelization of mass assignment and Fourier convolution}
A race condition may occur in mass assignment 
because it is possible for PEs to
have particles which write to the same elements of the mass array. The
approaches to solving this problem are numerous but consist mainly of two
ideas; (a) selectively assign particles to PEs so that mass assignment
occurs at grid cells that do not overlap, thus race condition
is avoided or (b) use ghost cells and contiguous slabs of 
particles which are constrained in their extent in the simulation space.
The final mass array must be accumulated by adding up all cells, 
including ghosts.
Ghost cells offer the advantage that they allow the calculation to be  
load-balanced (the size of a slab may be adjusted) but require more 
memory. Controlling which particles are
assigned
does not require more memory but may cause a load imbalance. Because the
types of simulation performed have particle distributions that can vary
greatly, both of these algorithms have been implemented.

\subsubsection{Using controlled particle assignment}\label{cpa}
The particles in the simulation are ordered in the z-direction within the
chaining cells. Because the chaining cells are themselves 
ordered along 
the z-axis (modulo their cubic arrangement) a naive solution would be to 
simply divide up the list 
of particles. However, this approach does not prevent a race condition 
occurring, it merely makes it less likely. In the CRAFT code the race 
condition was avoided by
using the `{\em{atomic update}}' facility which is a
lock{\em--}fetch{\em--}update{\em--}store{\em--}unlock hardware primitive
that allows fast updating of arrays where race conditions are present. 
Modern cache coherency protocols are unable to provide this kind of 
functionality.

Using the linked/ordered list to control the
particle assignment provides an elegant solution to the race condition 
problem. Since the linked list encodes the position of a
particle to within a chaining cell, it is possible to selectively assign
particles to the mass array that do not have overlapping writes.  To
assure a good load balance it is better to use columns ($Ls\times 
B\times
B$, where $Ls$ is the size of the chaining mesh and $B$ is a number of
chaining cells)  of cells rather than slabs ($Ls \times Ls \times B$).
Since there are more columns than slabs a finer grained distribution of 
the
computation can be achieved and thus a better load balance. 
This idea can also be
extended to a 3-d decomposition, however in simple
experiments we have found this approach to be inefficient for all but the
most clustered particle distributions (in particular cache reuse is
lowered by using a 3-d decomposition).

\begin{figure}[t]
\vspace{8cm}
\includegraphics{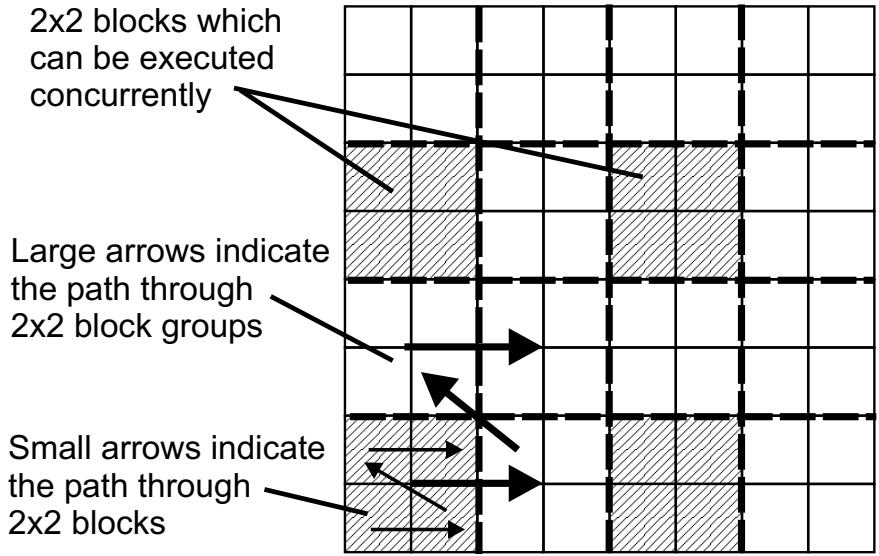}
\caption{Cell grouping and sorting in the $2\times2$ configuration
scheme.}
\label{2by2}
\end{figure}

Chaining mesh cells have a minimum width of 2.2 potential mesh cells in
\Hydra and \fig~\ref{chaining} displays a plot of the chaining mesh
overlaid on the potential mesh. When performing mass assignment for a
particle, writes will occur over all 27 grid cells found by the TSC
assignment scheme. Thus providing a buffer zone of one cell is not
sufficient to avoid the race condition since particles in chaining cells
one and three may still write to the same potential mesh cell. A spacing
of two chaining mesh cells is sufficient to ensure no possibility of
concurrent writes to the same mesh cell. The ``buffer zones'' thus 
divide
up the simulation volume into a number of regions that can calculated
concurrently and those that cannot.  Moreover, there will be need to be
a series of barrier synchronizations as regions that can be written
concurrently are finished before beginning the next set of regions. The
size of the buffer zone means that there are two distinct ways of
performing the mass assignment using columns: \begin{itemize}

\item $Ls\times 1 \times 1$ columns in $3\times 3$ groups. 
Assign
mass for particles in each of the columns simultaneously and then 
perform a
barrier synchronization at the end of each column. Since the columns are 
in
$3\times3$ groups there are nine barriers.

\item $Ls\times 2\times 2$ columns which are grouped into $2\times 
2$
groups. In this case the number of barriers is reduced to four, and if 
desired, 
the size of the column can be increased beyond two while still 
maintaining four barriers. 
However, load-imbalance under clustering argues against this idea. 

\end{itemize}
 See \fig~\ref{2by2} for a graphical representation of the algorithm.

To improve load balance, a list of the relative work in each column
(that can be evaluated before the barrier synchronization) is calculated
by summing over the number of particles in the column.  Once the
workload of each column has been evaluated, the list of relative
workloads is then sorted in descending order. The calculation then
proceeds by dynamically assigning the list of columns to the PEs as they
become free. The only load imbalance then possible is a wait for the
last PE to finish which should be a column with a low workload. Static,
and even cyclic, distributions offer the possibility of more severe load
imbalance.

For portability reasons, we have parallelized the FFT by hand rather than 
relying on a threaded 
library such as provided by FFTW. The 3-d FFT is 
parallelized over 
`lines' 
by calling a series of 1-d FFTs. We perform the transpose 
operation by explicitly copying contiguous pieces of the main data array 
into buffers which have a long stride. This improves data locality of the 
code considerably as the 
stride has been introduced into the buffer which is a local array. The 
FFTs are then performed on the buffer, and values are finally copied back 
into the data arrays. 
The convolution which follows the FFT relies upon another set of nested 
loops in the axis directions. To enable maximum granularity we have 
combined 
the z- and y-directions into one larger loop which is then statically 
decomposed among the processors.
Parallel efficiency is high for this method since if the number of 
processors divides the size of the FFT grid we have performed a 
simple slab decomposition of the serial calculation. 

\subsection{Parallelization of the PP and SPH force components}
The short range forces are accumulated by using 3 nested loops to sort
through the chaining mesh. As in mass assignment, a race condition is 
present due to the possibility of concurrent writes to the data arrays.
Again, in the CRAFT code, this race condition was avoided
by using the atomic update primitive.  

Because a particle in a given chaining mesh cell may write to its 26 
nearest-neighbour cells it is necessary to provide a two cell 
buffer 
zone. We can therefore borrow the exact same column decomposition that 
was used in mass assignment.
Tests
showed that of the two possible column sorting algorithms discussed in
section~\ref{cpa}, $Ls\times 2\times 2$ columns are more efficient than
the
$Ls\times 1\times1$ columns. The difference in execution time in
unclustered
states was negligible, but for highly clustered distributions (as measured
in the Santa Barbara cluster simulation \cite{SB99}), the $Ls\times 
2\times 
2$ method 
was
approximately 20\% faster. This performance improvement is attributable 
to
the difference in the number of barrier synchronizations required by each
algorithm (four versus nine) and also the better cache reuse of the 
$Ls\times 2\times 2$ columns.

\subsection{Task farm of refinements} 
As discussed earlier, the smaller sub-meshes ($N_r\leq262,144$) are 
distributed as a
task farm amongst the PEs. As soon as one processor becomes free it is
immediately given work from a pool via the dynamic scheduling option in 
OpenMP. Load imbalance may still occur in the task
farm if one refinement takes significantly longer than the rest and there
are not enough refinements to balance the workload over the remaining PEs.
Note 
also the task farm is divided into levels, the refinements placed within
the top level, termed `level one refinements' must be completed before
calculating the `level two refinements', that have been
generated by the level one refinements. 
However, we 
minimize the impact of the barrier wait by sorting refinements by the
number of particles contained within them and then begin calculating the
largest refinements first.  This issue emphasizes one of the drawbacks
of a shared memory code\-it is limited by the parallelism available and
one has to choose between distributing the workload over the whole machine or
single CPUs. It is not possible in the OpenMP programming environment to
partition the machine into processor groups. This is the major drawback
that has been addressed by the development of an MPI version of the code 
\cite{T03}.

\section{Considerations for NUMA architectures}
Because of the comparatively low ratio of work to memory read/write 
operations the code is potentially sensitive to memory latency issues.   
To test this sensitivity in a broad sense, we have examined
the performance of the code for a range of problem sizes, from
$2\times16^3$ particles to $2\times128^3$, the smallest of which
is close to fitting in L2 cache. A strong latency dependence will
translate into much higher performance for problem sizes resident in
cache as opposed to those requiring large amounts of main memory.
We also consider the performance for both clustered and unclustered
particle distributions since the performance envelope is considerably
different for these two cases. The best metric for performance is
particle updates per second, since for the unclustered distribution
P${}^3$M has an
operation dependence dominated by ${\bf O}(N)$ factors, while in the
clustered state the algorithm dominated by the cost of the SPH
solution which also scales as ${\bf O}(N)$.

\begin{figure}[t]
\vspace{10cm}
\includegraphics{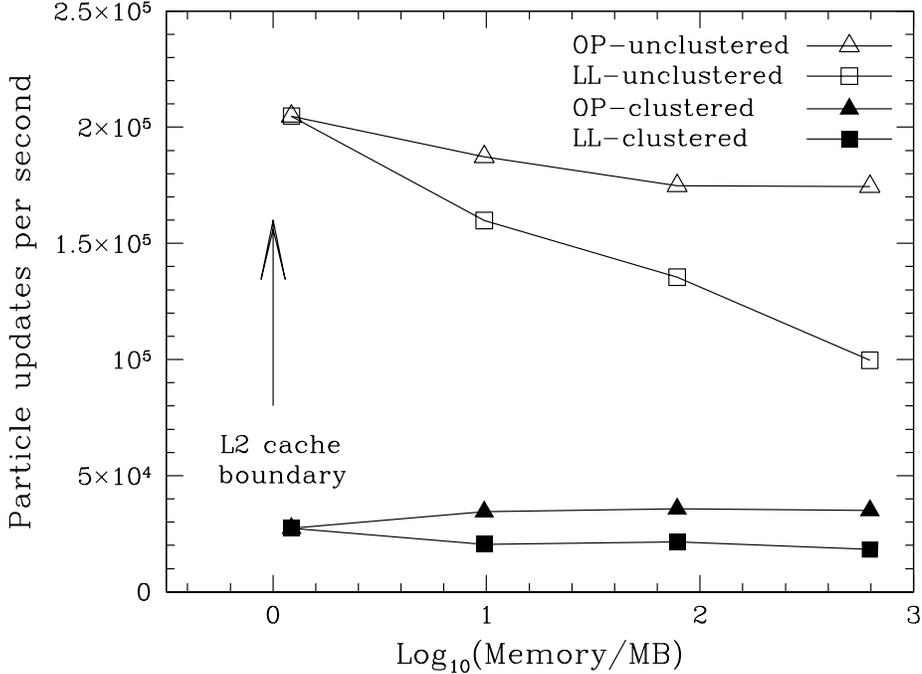}
\caption{Performance of the code for clustered and unclustered
distributions of
sizes between
$2\times16^3$ and $2\times128^3$, on a 2Ghz AMD Opteron.
The logarithm of memory consumption
in megabytes, rather
than particle
number, is
plotted along the x-axis. Both the ordered particle (OP) and linked list
(LL) versions
of the code were used.
The clustered state exhibits a similar level
of RMS clustering to the Santa Barbara simulation discussed in section
6. Comparatively little dependence upon latency is observed.}
\label{latency}
\end{figure}

The results are plotted in figure \ref{latency}, as a function of memory
consumption. We find that the $2\times16^3$
simulations show equal performance for both the linked list and ordered
particle code under both clustering states. However, for larger problem
sizes the unclustered state shows a considerable drop-off in performance
for the linked list code, while the ordered particle code begins to
level off at the $2\times64^3$ problem size. The clustered distributions
show little sensitivity to problem size, which is clearly indicative of
good cache reuse and a lack of latency sensitivity. We conclude that the
algorithm is comparatively insensitive to latency because the solution
time is dominated largely by the PP part of the code which exhibits good
cache reuse.

The increased performance improvement seen for the ordered particle
code is caused by the increased data locality. On NUMA architectures
this has a direct benefit as although the 
penalty for distant memory fetches is large (several hundreds of 
nanoseconds) the cache reuse ensures this penalty is only felt rarely.
We have found that the
locality is sufficiently high to render direct data placement largely
irrelevant on the SGI Origin. The only explicit data placement
we perform is a block distribution of the particle data over PEs. The 
constant reordering of
particles ensures that this is an effective distribution. For the
remainder of the arrays we use the ``first touch'' placement paradigm,
namely that the first PE to request a specific  memory page is assigned
it. Despite the simplicity, this scheme works very effectively.

Since the granularity of the chaining cells is smaller than the smallest
memory page size, prefetching is better strategy than memory page 
rearrangement.
This works particularly
effectively in the PP part of the algorithm where a comparatively large
amount of work is done per particle. In this section of code we specify
that two cache lines should always be retrieved for each cache
miss, and we also
allow the compiler to make further (aggressive) prefetching
predictions. The net effect
of this is to almost completely hide the latency on the Origin. This
can be seen in the performance scaling, where excellent results are  
achieved up to 64 nodes (see section~\ref{perf}).

However, there is one particularly noticeable drawback to NUMA
architectures. A number of the arrays used within the PM solver are
equivalenced to a scratch work space within a common block. First touch
placement means that the pages of the scratch array are distributed
according to the layout of the first array equivalenced to the common
block. If the layout of this array is not commensurate with the layout of
subsequent arrays that are equivalenced to the scratch area then severe
performance penalties result. Our solution has simply been to remove the
scratch work space and suffer the penalty of increased memory
requirements. 

\section{Performance}\label{perf}
\subsection{Correctness Checking}
Our initial tests of correctness of large simulations 
($2\times256^3$), comparing serial to parallel 
runs, showed variation in global values, such as the total mass within the 
box at the 0.01 percent level. However, this turned out to be a precision 
issue, as 
increasing the summation variables to double precision removed any 
variation in values. With these changes made, 
we have confirmed that the parallel code gives identical results to the 
serial code to machine-level rounding errors. An extensive suite 
of 
tests of the \Hydra code are detailed in \cite{CT95} and \cite{rob2}.

\subsection{Overall Speed}

Our standard test case for benchmarking is the `Santa Barbara cluster'
used in the paper by Frenk \etal \cite{SB99}. This simulation models the
formation of a galaxy cluster of mass $1.1\times 10^{15}$ \msol  in a
Einstein-de Sitter sCDM cosmology with parameters $\Omega_d=0.9$,
$\Omega_b$=0.1, $\sigma_8$=0.6, $H_0=0.5$, and box size 64 Mpc.  Our 
base
simulation cube has $2\times64^3$ particles, which yields 15300 particles 
in
the galaxy cluster, and we use an S2 softening length of 37 kpc. 
Particle masses are $6.25\times10^{10}$ \msol  for dark matter and 
$6.94\times10^9$ \msol  for gas. To prepare a
larger data set we simply tile the cube as many times as necessary.
An output from z=7.9 is used as an `unclustered' data set, and one from 
z=0.001 as a
`clustered' data set.

We were given access to two large SMP machines to test our code on, a 64 
processor SGI Origin 3000 (O3k, hereafter) at the 
University of 
Alberta and a 64 processor
Hewlett Packard GS1280 Alphaserver. Both of these machines have NUMA 
architectures, the O3k topology being a hypercube, while the GS1280 uses a 
two dimensional torus. The processors in the O3k are 400 Mhz MIPS R12000 
(baseline SPECfp2000 319)
while the GS1280 processors are 21364 EV7 Alpha CPUs running at 1150 Mhz
(baseline SPECfp2000 1124). There is an expected raw performance 
difference of over a factor of three between the two CPUs, although in 
practice we find the raw performance difference to be slightly over two.

\begin{figure}[t]
\vspace{6cm}
\includegraphics{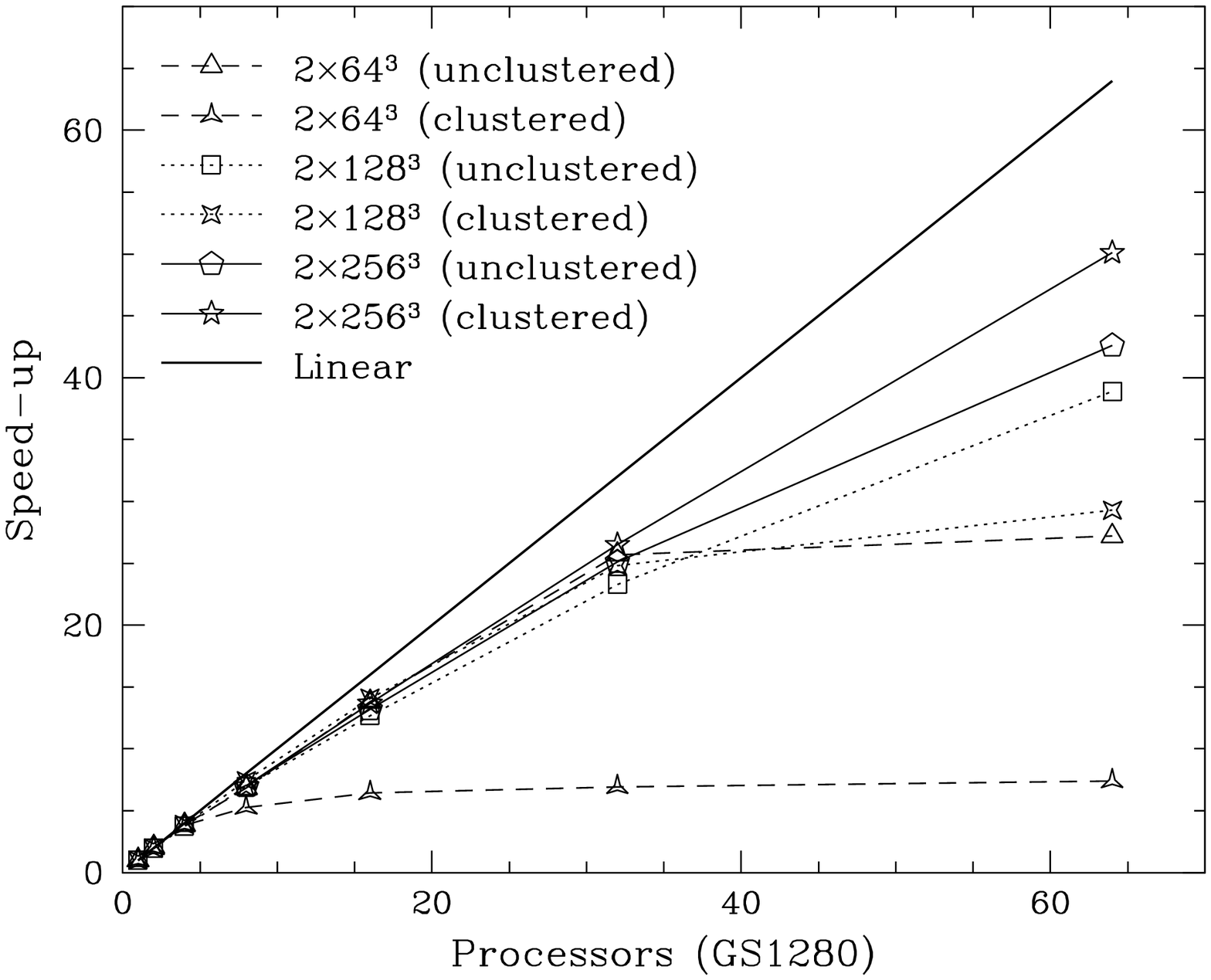}
\includegraphics{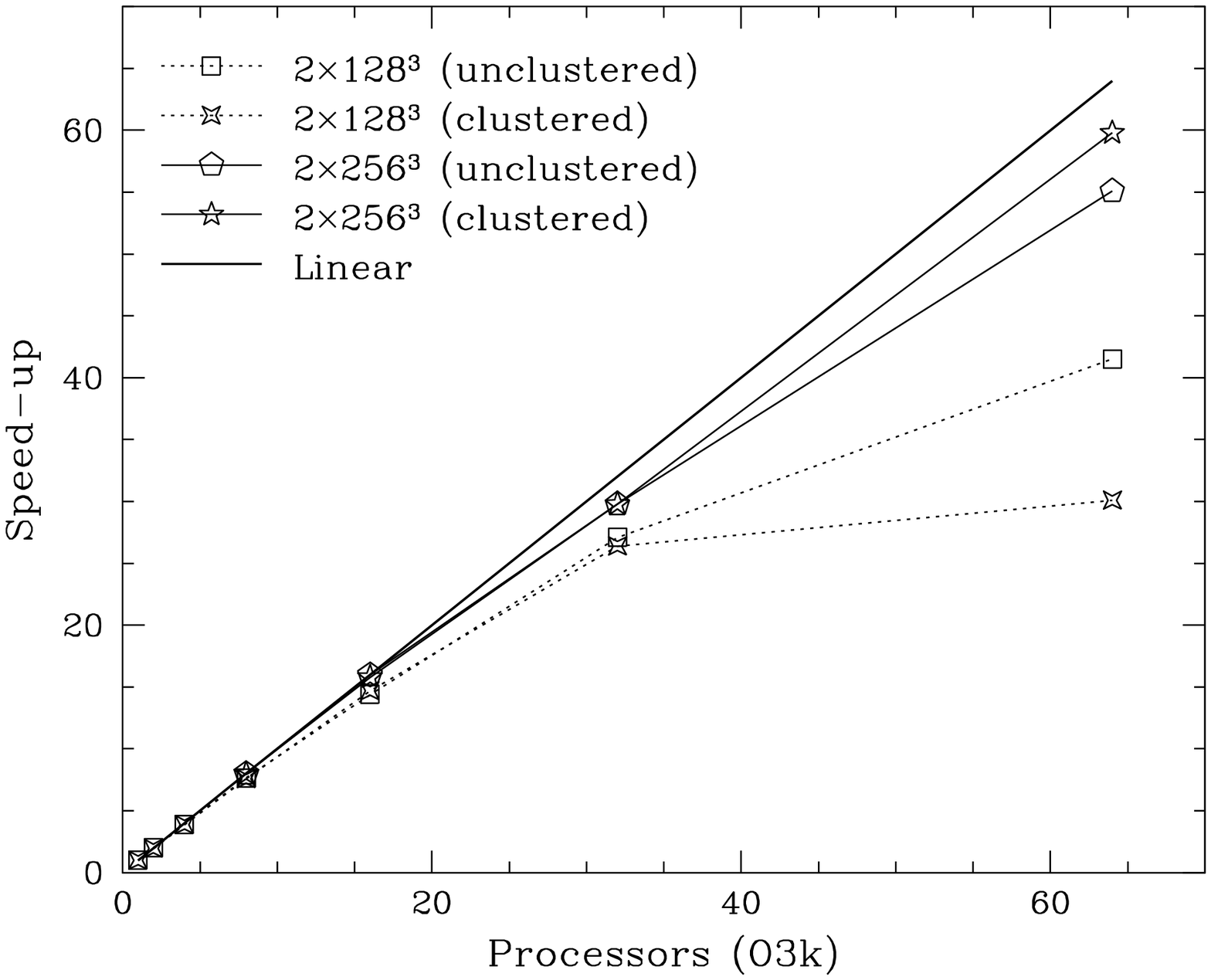}
\caption{
Parallel speed-up for various particle configurations and processor 
counts in the strong scaling regime for the GS1280 and O3k. Open polygons 
correspond to the z=7.9 
dataset, pointed stars to 
the z=0.001 data set. Dashed lines correspond to the $2\times64^3$ data, 
dotted lines to the $2\times128^3$ and the thin solid line to the 
$2\times256^3$ data. Perfect linear scaling is given by the thick solid 
line. Provided there is sufficient parallel work available, scaling is 
excellent to 32 processors (only the clustered $2\times64^3$ run exhibits 
a notable lack of scaling). Beyond 32 processors only the $2\times256^3$ 
runs have sufficient work to scale well.} 
\label{scaling} 
\end{figure}

We conducted various runs with 
differing particle and data sizes to test scaling in both the strong 
(fixed problem size) and weak (scaled problem size) regimes.
The parallel speed-up and raw execution times are 
summarized in tables \ref{tab1} \& \ref{tab2} and speed-up is shown 
graphically in 
figure \ref{scaling}. 
Overheads associated with I/O and start-up are not included. Further, we 
also do not include the overhead associated with placing refinements on 
the top level of the simulation, as this is only performed every 20 
steps.

With the exception of the clustered $2\times64^3$ run, parallel scaling is
good (better than 73\%) to 32 processors on both machines for all runs. 
The clustered $2\times64^3$
simulation does not scale effectively because the domain decomposition is 
not sufficiently fine to deal with the load imbalance produced by  
this particle configuration.
Only the largest simulation has sufficient work to scale effectively 
beyond 32 processors. 
 To estimate the scaling of the $2\times256^3$ 
run 
we estimated the speed-up on 8 nodes of the GS1280 as 7.9 (based upon the 
slightly lower efficiencies observed on the $2\times128^3$ compared to 
the O3k), while on the 
O3k we estimated the speed up as 8.0. We then estimated the scaling 
from that point. Speed-up relative to the 8 processor value is also given 
in table 1, and thus values may be scaled as desired.

\renewcommand{\arraystretch}{1.0}
\begin{table}
\caption{Parallel scaling efficiencies and wall 
clock timings for a full
gravity-hydrodynamic calculation on the SGI Origin 3000. Results in 
parenthesis indicate that
the values are estimated. The 64 processor results for the two smallest 
runs have been omitted because they resulted in a slowdown relative to 
the 32 processor run.} 
\begin{center}
\begin{tabular}{c l c c c c c}
\hline
N & Mesh & PEs & Redshift & Wall Clock/s & Speed-up & Efficiency\\
\hline
$2\times 64^3$ & $128^3$& 1 & 7.9  & 12.2 & 1.00 & 100\% \\
 $2\times 64^3$ & $128^3$& 2 & 7.9 &  6.27 & 1.95 & 98\% \\
 $2\times 64^3$ & $128^3$& 4 & 7.9  & 3.27 & 3.73 & 93\% \\
 $2\times 64^3$ & $128^3$& 8 & 7.9  & 1.70 & 7.18 & 90\% \\
 $2\times 64^3$ & $128^3$& 16 & 7.9 & 0.94 & 13.0 & 81\% \\
 $2\times 64^3$ & $128^3$& 32 & 7.9 & 0.60 & 20.3 & 63\% \\
 &  &  & & & & \\
$2\times 64^3$ & $128^3$& 1 & 0.001  & 53.9 & 1.00 & 100\% \\
 $2\times 64^3$ & $128^3$& 2 & 0.001  & 27.2 & 1.98 & 99\% \\
 $2\times 64^3$ & $128^3$& 4 & 0.001  & 14.0  & 3.85 & 88\% \\
 $2\times 64^3$ & $128^3$& 8 & 0.001  & 9.27  & 5.81 & 73\% \\
 $2\times 64^3$ & $128^3$& 16 & 0.001 & 8.16  & 6.61 & 41\% \\
 $2\times 64^3$ & $128^3$& 32 & 0.001 & 8.10  & 6.65 & 21\% \\
 &  &  & & & & \\
  $2\times 128^3$ & $256^3$&1 & 7.9  & 105  & 1.00 & 100\% \\
  $2\times 128^3$ & $256^3$&2 & 7.9  & 51.9 & 2.02 & 101\% \\
  $2\times 128^3$ & $256^3$&4 & 7.9  & 26.8 & 3.91 & 98\% \\
  $2\times 128^3$ & $256^3$&8 & 7.9  & 13.7 & 7.66 & 96\% \\
  $2\times 128^3$ & $256^3$&16 & 7.9 &  7.28 & 14.4 & 90\% \\
  $2\times 128^3$ & $256^3$&32 & 7.9 &  3.88 & 27.1 & 85\% \\
  $2\times 128^3$ & $256^3$&64 & 7.9 &  2.53 & 41.5 & 65\% \\
 &  &  & & & & \\
  $2\times 128^3$ & $256^3$&1 & 0.001  & 407   & 1.00 & 100\% \\
  $2\times 128^3$ & $256^3$&2 & 0.001  & 208 & 1.96 & 98\% \\
  $2\times 128^3$ & $256^3$&4 & 0.001  & 105 & 3.88 & 97\% \\
 $2\times 128^3$ & $256^3$&8 & 0.001  &  53.6 & 7.59 & 95\% \\
 $2\times 128^3$ & $256^3$&16 & 0.001 &  27.6 & 14.7 & 92\% \\
  $2\times 128^3$ & $256^3$&32 & 0.001 & 15.4 & 26.4 & 83\% \\
  $2\times 128^3$ & $256^3$&64 & 0.001 & 13.5 & 30.1 & 47\% \\
 &  &  & & & & \\
 $2\times 256^3$ & $512^3$& 8 & 7.9 & 115.   & (8.0) & (100\%)\\
 $2\times 256^3$ & $512^3$& 16 & 7.9 & 57.5  & (16.0)[2.00] & (100\%)\\
 $2\times 256^3$ & $512^3$& 32 & 7.9 & 30.9  & (29.8)[3.72] & (93\%)\\
 $2\times 256^3$ & $512^3$& 64 & 7.9 & 16.7  & (55.1)[6.89] & (86\%)\\
 &  &  & & & & \\
 $2\times 256^3$ & $512^3$& 8 & 0.001 & 484   & (8.0) & (100\%)\\
 $2\times 256^3$ & $512^3$& 16 & 0.001 & 245  & (15.8)[1.98] & (100\%)\\
 $2\times 256^3$ & $512^3$& 32 & 0.001 & 130 & (29.8)[3.72] & (93\%)\\
  $2\times 256^3$ & $512^3$& 64 & 0.001 & 64.7 & (59.8)[7.48] & (93\%)\\
\hline
\end{tabular}
\end{center}
                                                                                
\vspace*{.6cm}
\noindent
\label{tab1}
\end{table}

\linespread{0.75}
\begin{table}
\caption{Parallel scaling efficiencies and wall clock timings for a full
gravity-hydrodynamic calculation calculation on the HP GS1280. 
Results in parenthesis indicate that 
the values are estimated.} 
 \begin{center}
\begin{tabular}{c l c c c c c}
\hline
N & Mesh & PEs & Redshift & Wall Clock/s & Speed-up & Efficiency\\
\hline
$2\times 64^3$ & $128^3$& 1 & 7.9  & 5.13 & 1.00 & 100\% \\
 $2\times 64^3$ & $128^3$& 2 & 7.9 &  2.50 & 2.06 & 103\% \\
 $2\times 64^3$ & $128^3$& 4 & 7.9  & 1.33 & 3.86 & 97\% \\ 
 $2\times 64^3$ & $128^3$& 8 & 7.9  & 0.75 & 6.84 & 86\% \\ 
 $2\times 64^3$ & $128^3$& 16 & 7.9 & 0.37 & 13.8 & 86\% \\ 
 $2\times 64^3$ & $128^3$& 32 & 7.9 & 0.20 & 25.7 & 80\% \\ 
 $2\times 64^3$ & $128^3$& 64 & 7.9 & 0.19 & 27.2 & 43\% \\
 &  &  & & & & \\
 $2\times 64^3$ & $128^3$& 1 & 0.001  & 20.7 & 1.00 & 100\% \\
 $2\times 64^3$ & $128^3$& 2 & 0.001  & 10.5 & 1.98 & 99\% \\ 
 $2\times 64^3$ & $128^3$& 4 & 0.001  &  5.38 & 3.84 & 96\% \\
 $2\times 64^3$ & $128^3$& 8 & 0.001  &  3.94 & 5.25 & 67\% \\
 $2\times 64^3$ & $128^3$& 16 & 0.001 &  3.21 & 6.45 & 40\% \\
 $2\times 64^3$ & $128^3$& 32 & 0.001 &  2.99 & 6.92 & 22\% \\
$2\times 64^3$ & $128^3$& 64 & 0.001 &  2.80 & 7.39 & 12\% \\
 &  &  & & & & \\
  $2\times 128^3$ & $256^3$&1 & 7.9  & 41.2  & 1.00 & 100\% \\
  $2\times 128^3$ & $256^3$&2 & 7.9  & 21.0 & 1.96 & 98\% \\  
  $2\times 128^3$ & $256^3$&4 & 7.9  & 11.0 & 3.75 & 94\% \\  
  $2\times 128^3$ & $256^3$&8 & 7.9  &  5.92 & 6.96 & 87\% \\ 
  $2\times 128^3$ & $256^3$&16 & 7.9 &  3.26 & 12.7 & 79\% \\ 
  $2\times 128^3$ & $256^3$&32 & 7.9 &  1.77 & 23.3 & 73\% \\
  $2\times 128^3$ & $256^3$&64 & 7.9 &  1.06 & 38.9 & 61\% \\ 
 &  &  & & & & \\
  $2\times 128^3$ & $256^3$&1 & 0.001  & 154   & 1.00 & 100\% \\
  $2\times 128^3$ & $256^3$&2 & 0.001  &  77.7 & 1.98 & 99\% \\ 
  $2\times 128^3$ & $256^3$&4 & 0.001  &  39.7 & 3.88 & 97\% \\ 
 $2\times 128^3$ & $256^3$&8 & 0.001  &   20.7 & 7.44 & 93\% \\ 
 $2\times 128^3$ & $256^3$&16 & 0.001 &   10.9 & 14.1 & 88\% \\ 
  $2\times 128^3$ & $256^3$&32 & 0.001 &   6.2 & 24.8 & 76\% \\
  $2\times 128^3$ & $256^3$&64 & 0.001 &   5.3 & 29.3 & 46\% \\ 
 &  &  & & & & \\
 $2\times 256^3$ & $512^3$& 8 & 7.9 & 49.5   & (7.9) & (99\%)\\
 $2\times 256^3$ & $512^3$& 16 & 7.9 & 26.4  & (14.9)[1.88] & (93\%)\\
 $2\times 256^3$ & $512^3$& 32 & 7.9 & 13.8  & (28.4)[3.59] & (89\%)\\
 $2\times 256^3$ & $512^3$& 64 & 7.9 & 8.13  & (48.1)[6.09] & (75\%)\\
 &  &  & & & & \\
 $2\times 256^3$ & $512^3$& 8 & 0.001 & 215   & (7.9) & (99\%)\\
 $2\times 256^3$ & $512^3$& 16 & 0.001 & 110  & (15.4)[1.95] & (96\%)\\
 $2\times 256^3$ & $512^3$& 32 & 0.001 & 56.7 & (29.9)[3.79] & (93\%)\\
  $2\times 256^3$ & $512^3$& 64 & 0.001 & 30.0 & (56.6)[7.16] & (88\%)\\
\hline
\end{tabular}
\end{center} 

\vspace*{.6cm}
\noindent
\label{tab2}
\end{table}
\linespread{1.0}

To quantify our results further we summarize the performance of the code
using a popular performance metric for cosmological codes, namely the
number of particle updates per second. As a function of 
the number of
nodes within the calculation this also gives a clear picture of the
scaling achieved. 
Because the simulation results we obtained were run using the combined 
gravity-hydrodynamic solver it is necessary for us to interpolate the 
gravitational speed. To do this we calculated the ratio of the code speed 
with and without hydrodynamics, and also without the PP correction, on 1 
CPU of our local GS160 Alphaserver, 
and on 1 CPU of the O3k. To 
ensure 
this approximation is as reasonable as possible we calculated the ratios 
for both the z=7.9 and z=0.001 datasets. Relative to the speed obtained 
for the combined solver, the gravity-only solver was found to be 
1.63(1.29) 
times faster for the z=7.9 dataset and 1.84(1.49) times faster for the 
z=0.001 
dataset, for the GS1280 (and O3k). The PM speed was found to 2.4(2.5) 
times 
faster for the z=7.9 dataset and 9.21(10.3) times faster for the
z=0.001 
dataset. 

\begin{figure}[t]
\vspace{6cm}
\includegraphics{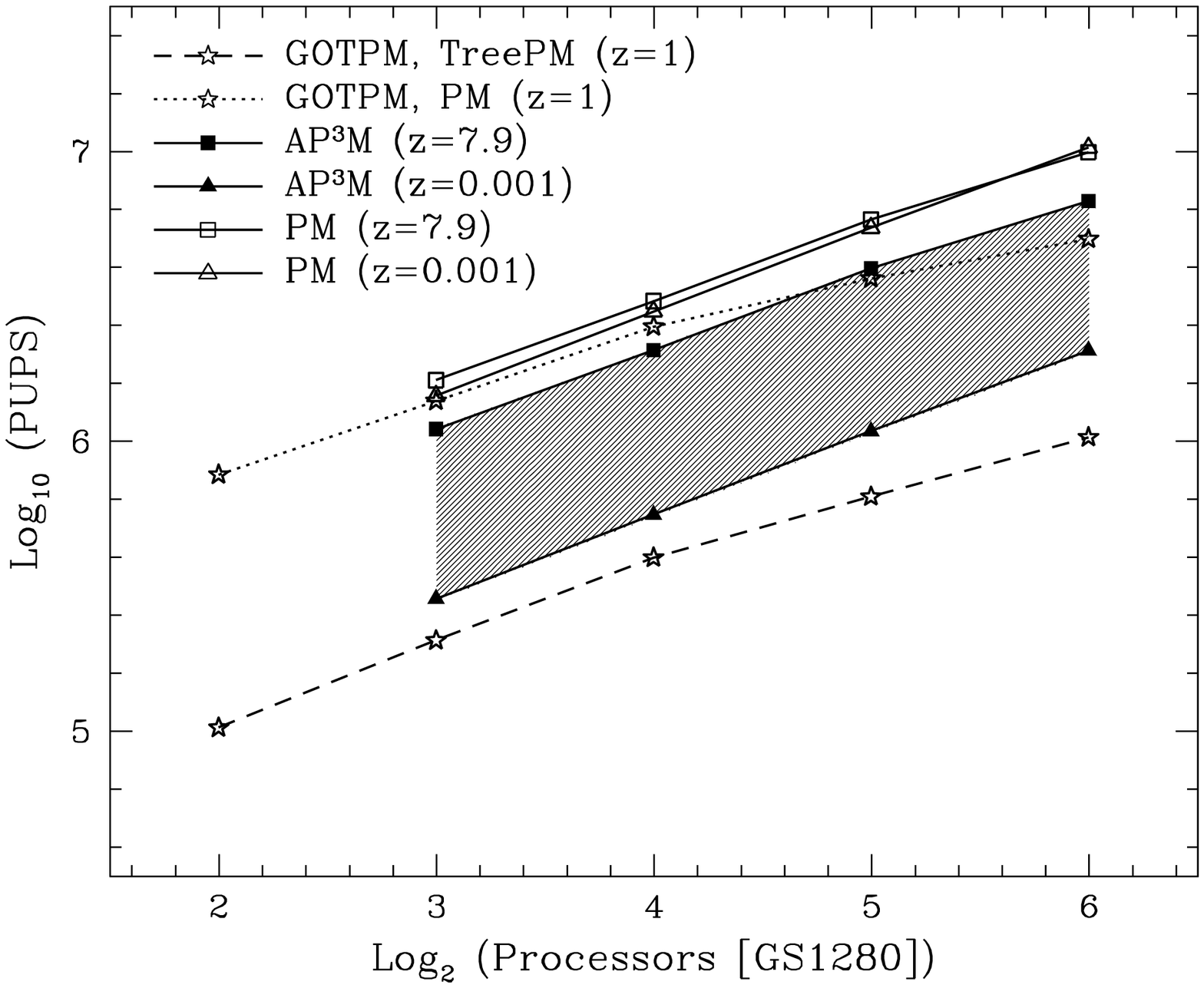}
\includegraphics{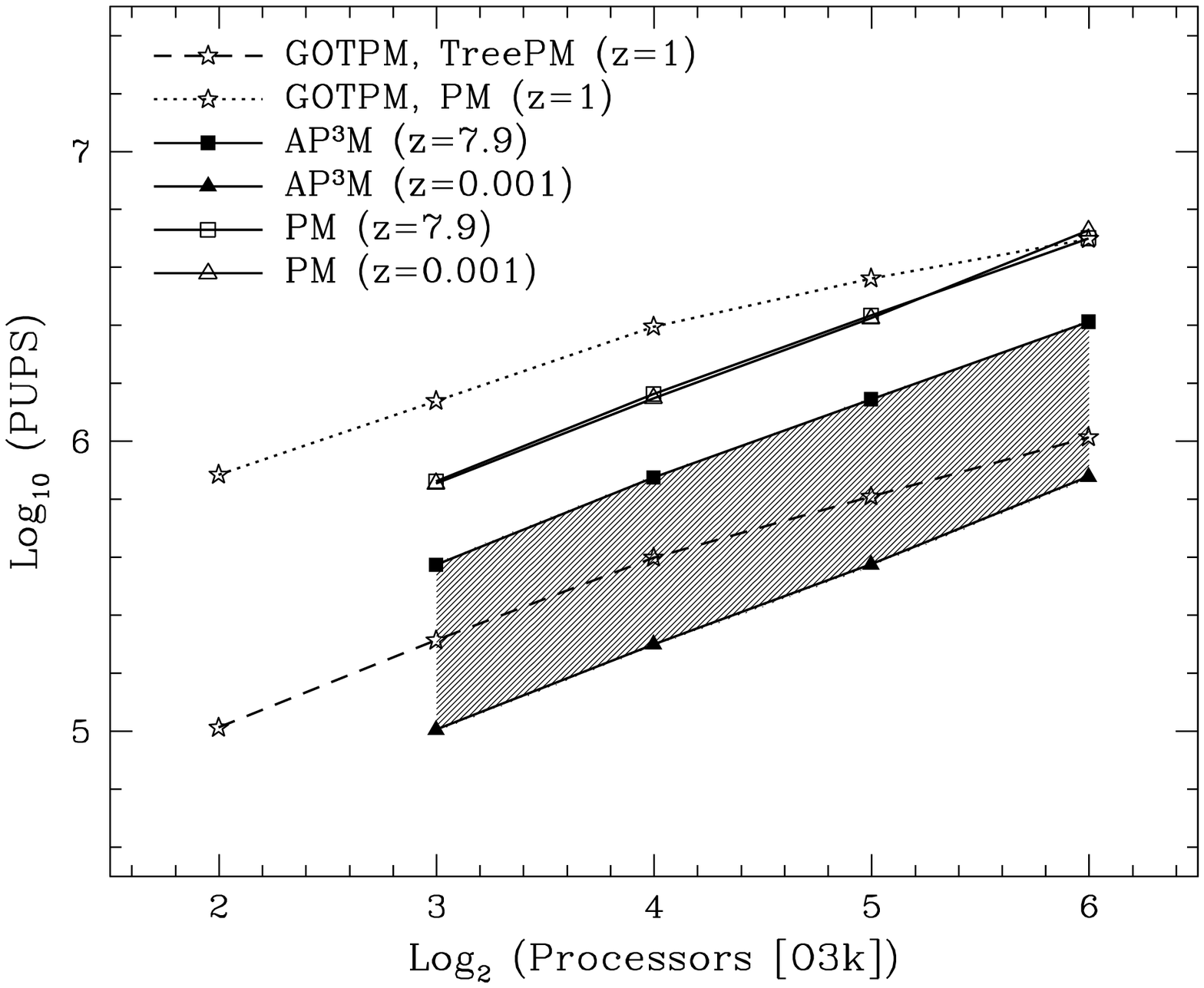}  
\caption{Performance of the gravitational solver measured by particle 
updates per second (PUPS) on the GS1280 and O3k. Values are given for both 
the raw 
PM speed as well as the full \ap3m solution, with the shaded area denoting 
the performance region for the code from unclustered to clustered 
distributions. For comparison,  data
for the Tree-PM (``GOTPM'') code of Dubinski et al for a data-set with
comparable mass resolution, at an expansion epoch of z=1, are provided.
Note that the 
GOTPM figures are for a less clustered distribution than our z=0.001 
dataset, however the processors used were approximately 14\% slower than those 
of the GS1280, and they used a high bandwidth Gigabit ethernet 
interconnect. Note that even the PM algorithms are not truly comparable 
since HYDRA uses a ten point difference for forces compared to the four 
point difference used in GOTPM. } \label{pups} \end{figure}

In figure \ref{pups} we show the estimated number of gravitational updates
per second achieved on in both the clustered and unclustered state of the 
$2\times128^3$ simulation (other simulation sizes show almost identical 
speeds) on the GS1280. The 
clustered state is approximately three times slower than the unclustered 
state for all simulation sizes. To provide comparison to other published 
work we have also included 
results presented by Dubinski \etal for a $256^3$ simulation conducted 
on 
a $512^3$ grid using a distributed memory Tree-PM code (``GOTPM''). 
Although a direct comparison of speed is not as instructive as might be 
hoped, since 
both the machine specifications and particle 
distributions differ, it is intriguing that the raw PM speed of both codes 
are very similar, with our code showing a moderate speed advantage 
(between 2.4 and 1.8 times faster depending on clustering).
Comparing the speed of the full solutions (for the $2\times256^3$ 
simulation) in the clustered state shows HYDRA to be 2.3 times faster, although 
the initial configuration is 3.9 times faster, while reportedly Tree-PM 
codes have a roughly constant cycle time with clustering \cite{B02}.
This highlights the fact that while Tree-PM codes have a roughly 
constant cycle time with clustering, there is still significant 
room for improving the execution on unclustered data sets.
It 
is also worth noting that,
as yet, our implementation of \ap3m lacks any multiple time-step 
capability, and implementing a 
mechanism that steps refinements within different time bins has 
potentially very significant performance gains. Such an integrator would 
bear similarities to the mixed-variable symplectic integrators used in 
planetary integrations \cite{WH92}. 

\subsection{Timing breakdown}
Although overall performance is the most useful measure of utility for 
the code, analysis of the time spent in certain code sections may elucidate 
performance bottlenecks.
Hence, for timing purposes, we break the code into three main sections; the 
top level PM, the top level PP and the refinement farm. The speed of list 
making and particle book-keeping is incorporated within these sections.

The execution time is initially dominated
by the solution time for the top level grid, but the growth of clustering makes the 
solution time strongly dependent upon the efficiency of the refinement farm.
While the top level solution
(necessarily) involves a large number of global barriers, the refinement
farm only uses a small number and performs a large number of independent
operations. The only exception is a critical section where the global list
of refinements is updated, however we ensure 
the critical section is only entered if a refinement has indeed derived 
new refinements. Thus, potentially, the refinement farm can scale better than the 
top level solution.

\begin{figure}[t]
\vspace{6cm}
\includegraphics{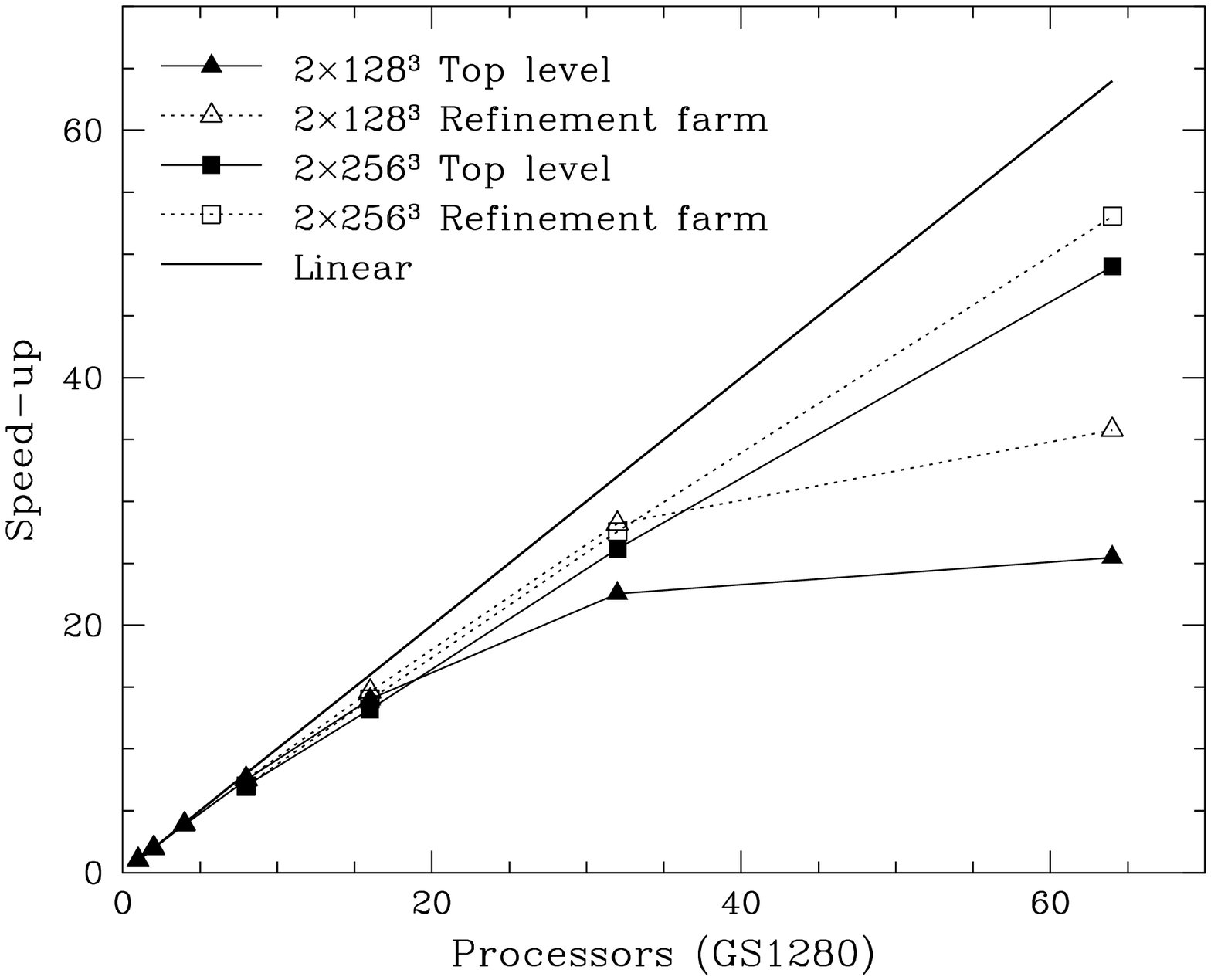}
\includegraphics{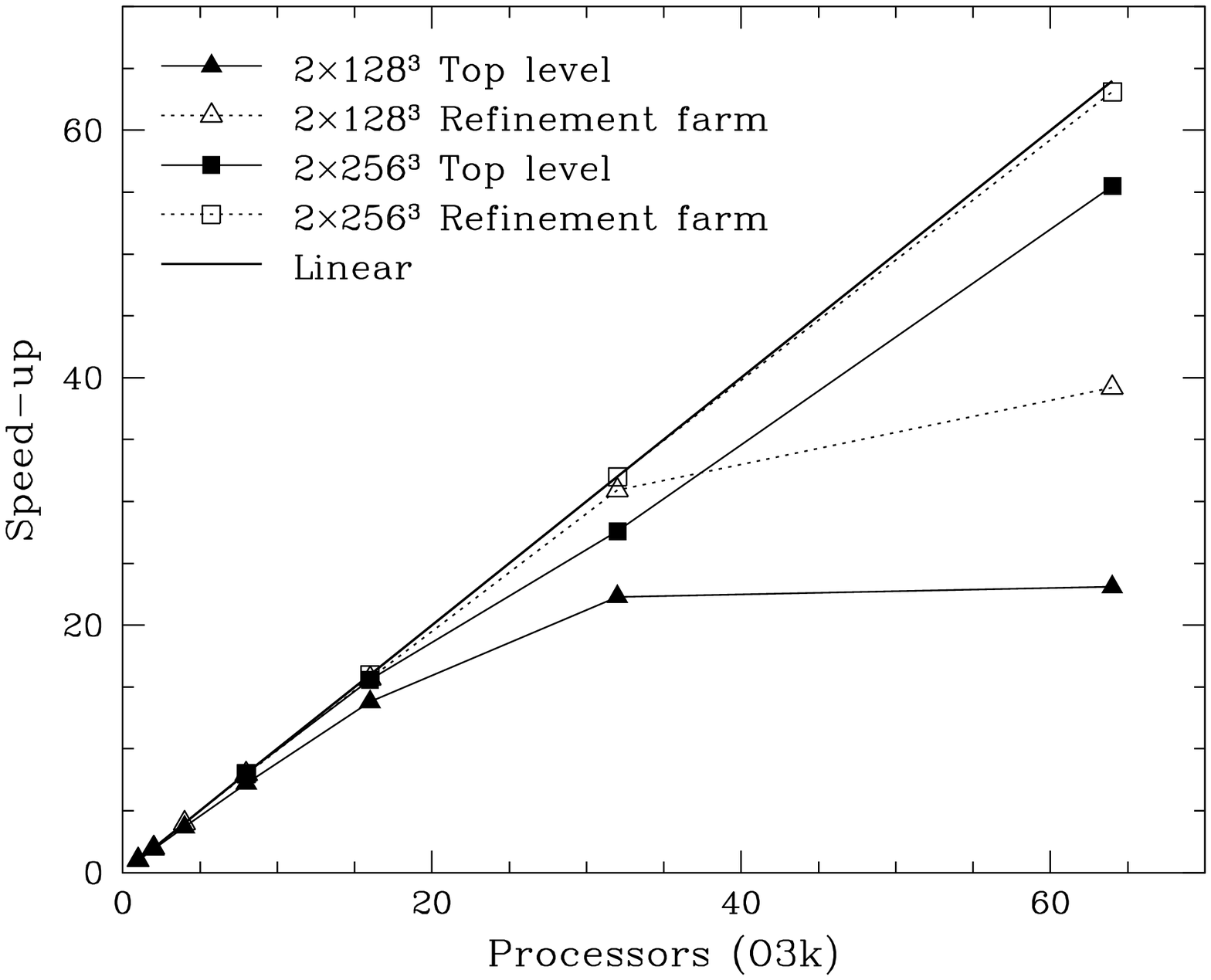}
\caption{Comparison of
parallel speed-up for the refinement farm versus the top level solver
for
different particle and processor counts. Scaling of the refinement farm on 
the 
O3k is  better than the GS1280 in both cases, and is almost perfect for 
the largest run out to 64 processors. 
The refinement farm does not scale as well in the $2\times128^3$ run as 
there is insufficient parallel work to scale out to 64 processors, 
however scaling to 32 is excellent. 
}
\label{farmvstop}
\end{figure}

In figure \ref{farmvstop} we plot the relative   
scaling of the top level solution compared to the refinement farm for a   
several different particle numbers. Provided sufficient work is 
available for distribution, the refinement farm is
seen to scale extremely well, with parallel efficiencies of 99\% and 83\%
observed for the $2\times256^3$ data set on 64 processors for the O3k and 
GS1280 respectively.

\section{Summary and Discussion}
Conducting high resolution simulations of cosmological
structure formation necessitates the use of parallel
computing. Although distributed architectures provide an abundance of
cheap computing power, the programming model for distributed systems is
fundamentally complex. Shared memory simplifies parallel programming
greatly since the shared address space means that only the calculation
itself need be distributed across nodes. In this paper we have discussed a
code for parallel shared memory computers that exhibits only marginally
higher complexity than a serial version of the code and which also
exhibits excellent performance. Additional constructs for parallel
execution introduce only a small (10\%) penalty for running on
1 node compared to the serial code.

Although the code does have some problems with regards load balancing, in
particular a deficit in performance occurs when a refinement is too large
to be calculated as part of the task farm but is not large enough to be
efficient across the whole machine, these situations are comparatively
rare. The poor scaling of SPH under heavy clustering is the most
significant cause of load imbalance. In particular, if the heavy
calculational load is confined to one refinement that is part of the task
farm all threads will block until this refinement is completed. The most
satisfactory solution to this problem is to substitute an alternative algorithm for
the SPH in high density regions. We will present details of an 
algorithm that improves the SPH cycle time for high density regions 
elsewhere (Thacker \etal in prep).

Most of the performance limitations can be traced to applying a grid code
in a realm where it is not suitable. As has been emphasized before,
treecodes are particularly versatile, and can be applied to almost 
any particle
distribution. However, for periodic simulations they become inefficient
since Ewald's method must used to calculate periodic forces. FFT-based
grid methods calculate the periodic force implicitly, and exhibit
particularly high performance for homogeneous particle distributions under
light to medium clustering. Highly clustered (or highly
inhomogeneous) particle distributions are naturally tailored to the
multi-timestepping capability of treecodes. Although we see scope for
introducing a multi-time stepping version of \ap3m where sub-grids are
advanced in different time step bins it is unclear in details what efficiencies
could be gained. There are clearly parts of the algorithm, such as
mass assignment, that are unavoidably subject to load imbalances. We
expect that since the global grid update would be required infrequently
the global integrator can still be made efficient. An 
efficient 
implementation of multiple time-steps is the last area where an order of 
magnitude improvement in simulation time can be expected for this class of algorithm. 

In terms of raw performance, the code speed is high relative to 
the values given by Dubinski et al. On the GS1280 the full 
solution time for the unclustered distribution even exceeds that of the PM 
solution quoted for 
GOTPM on 64 processors. \ap3m has been criticized previously for 
exhibiting a 
cycle time that fluctuates depending upon the underlying level of 
clustering. The data we have presented here shows the range in speeds is 
comparatively small (a factor of 4). We would also argue that since the 
cost of the short range correction is so small at early times, this 
criticism is misplaced.
While recent implementations of Tree-PM have an approximately constant 
cycle time irrespective of clustering, the large search radius used in the 
tree correction leads to the tree part of the algorithm dominating 
execution time for all stages of the simulation. Conversely, only at the 
end of the simulation is this true for HYDRA. 

Arguments have also been presented that suggest the PM cycle introduces
spurious force errors that can only been corrected by using a long range
PP correction (out to 5 PM cells). It is certainly true that PM codes
implemented with the so called `Poor Man's Poisson solver' \cite{BR69},
and Cloud-in-cell interpolation do suffer from large ($\sim$50\%)
directional errors in the force around 2-3 grid spacings.
However, as has been shown, first 
by Eastwood (see \cite{He81} for references) and more recently by 
Couchman, a combination of higher order 
assignment functions, Q-minimized Green's functions, and directionally 
optimized differencing functions can reduce errors in the inter-particle 
forces to sub 0.3\% levels (RMS). Surprisingly, although CIC gives a 
smooth force law (as
compared to
NGP), it does not reduce the angular isotropy of the mesh force.
Indeed, in two dimensions, moving from 
CIC to TSC interpolation  reduces directional errors from 50\% to 6\% and
Q-minimization of the Green's function reduces the anisotropy to sub 0.5\% 
levels \cite{E74}.
Furthermore, the technique of 
interlacing 
can be used to improve the accuracy of the PM force still further, 
but the additional FFTs required for this method rapidly lead to diminished 
returns. 

To date we have used this code to simulate problems ranging from galaxy 
formation to large-scale clustering. As emphasized in the introduction, the 
simple programming model provided by OpenMP has enabled us to rapidly 
prototype new physics algorithms which in turn has lead to the code being 
applied across a diverse range of astrophysics. Developing new physics 
models with this code takes a matter of hours, rather than the days typical of MPI 
coding.

We plan to make a new version of the code, incorporating more streamlined 
data structures and minor algorithmic improvements, publically available 
in the near future.

\section{Acknowledgments}
We thank an anonymous referee for comments which improved the paper.
Runs on the GS1280 machine were performed on our behalf by Andrew Feld of 
Hewlett Packard. We thank John Kreatsoulas for arranging time for us on 
this machine. Figures 1, 2, 4 and 5 were prepared 
by Dr L. Campbell.
RJT is funded in part by a CITA National Fellowship. 
HMPC acknowledges the support of NSERC and the Canadian
Institute for Advanced Research. SHARCNET and WestGrid computing 
facilities were used 
during this research.

\end{document}